\documentclass[preprint,prd,showpacs]{revtex4}

\usepackage{amsfonts, amssymb, amsthm, graphicx}
\usepackage{subfig}
\usepackage{amsmath}
\usepackage{float}

\newtheorem{de}{Definition}

\renewcommand{\theequation} {\arabic{section}.\arabic{equation}}

\newcommand{\be}{\begin{equation}}

\newcommand{\ee}{\end{equation}}

\newcommand{\bea}{\begin{eqnarray}}

\newcommand{\eea}{\end{eqnarray}}

\newcounter{orange}
\renewcommand{\theorange}{\alph{orange}}

\usepackage{epsfig}
\usepackage{epstopdf}
\usepackage{amsfonts, amssymb, amsthm}

\begin{document}
\title{The Wentzel -- Kramers -- Brillouin approximation method  applied to the Wigner function}

\author{J. Tosiek}\email{tosiek@p.lodz.pl}
\affiliation{Institute of Physics, Technical University of {\L}\'{o}d\'{z}\\
W\'{o}lcza\'{n}ska 219, 90-924 {\L}\'{o}d\'{z}, Poland.}

\author{R. Cordero}\email{cordero@esfm.ipn.mx}
\affiliation{Departamento de F\'{\i}sica, Escuela Superior de
F\'{\i}sica y Matem\'aticas del IPN\\ Unidad Adolfo
L\'opez Mateos, Edificio 9, 07738, M\'exico D.F., M\'exico.}

\author{F. J. Turrubiates}\email{fturrub@esfm.ipn.mx}
\affiliation{Departamento de F\'{\i}sica, Escuela Superior de F\'{\i}sica
y Matem\'aticas del IPN\\ Unidad Adolfo
L\'opez Mateos, Edificio 9, 07738, M\'exico D.F., M\'exico.}

\date{\today}

\begin{abstract}
An adaptation of the WKB method in the deformation quantization formalism is presented with the aim
to obtain an approximate technique of solving the eigenvalue problem for energy in the phase space quantum approach.
A relationship between the phase $\sigma(\vec{r})$ of a wave function $\exp \left(\frac{i}{\hbar} \sigma(\vec{r}) \right)$ and its respective Wigner function is derived. Formulas to calculate the Wigner function of  a product  and of a superposition of wave functions are proposed. Properties of a Wigner function of interfering states are also investigated. Examples of this quasi -- classical approximation in deformation quantization are analysed. A strict form of the Wigner function for states represented by tempered generalised functions has been derived.  Wigner functions of unbound states in the Poeschl -- Teller potential have been found.
\end{abstract}


\maketitle

\section{Introduction}
The number of physical problems which can be analytically solved is indeed limited. In fact, due to complexity of real systems such strict solutions are by their idealisation only approximations of reality. Thus finding  effective and universal approximate methods  is of vital importance. In this paper we propose an iterative algorithm of solving the integral $*$ -- eigenvalue equation for a Hamilton function in the scheme of deformation quantization. Our construction is based on the Wentzel -- Kramers -- Brillouin approximation (WKB) developed in the Hilbert space formulation of quantum mechanics.

The Schroedinger equation can be solved in an exact way when the form of the potential is relatively simple. For other cases several approximated methods have been developed during the years. One of such techniques is the WKB approximation mentioned in the previous paragraph \cite{daw,lan,lib,fro,gal} and sometimes called the quasi--classical approximation. This method was introduced in 1926 and is suitable for quantum systems where the potential changes slowly in comparison to the de Broglie wavelength.

The WKB algorithm  works well for the position representation. The main difficulty in its application arises from the fact that  the domain of a solution of the Schrodinger equation must be divided in spatially separable regions. On each region one gets a piece of the  wave function and the problem is to put these pieces together to obtain a global solution over all the regions. This obstacle is overcome with special connection formulas.

On the other hand, the  phase space quantum description is an alternative approach  to the Hilbert space formalism of quantum mechanics. This scheme provides supplementary valuable information in addition to the usual quantum formulation that is carried out in just one representation (position or momentum). In this framework the coordinates and momenta are considered simultaneously, which gives a natural extension of the Hamiltonian construction to describe quantum systems. A complete review of this topic can be found in \cite{zachos,kim,blas}.

In the phase space formulation of quantum mechanics the  information about the system is obtained through a Wigner function, which plays a similar role as  the wave function in usual
quantum mechanics. However, by means of this function it is possible  to study the classical limit in a more transparent way. The Wigner function has been employed widely in quantum optics, condensed matter, nuclear and particle physics, etc. A comprehensive guide to these applications can be found in \cite{kim} and the references cited therein.

A general way to obtain the phase space quantum description of a system is through the deformation quantization formalism. Its main advantage is that it can be used to treat systems with arbitrary phase spaces. To obtain a Wigner function for an arbitrary system it is necessary to find  solutions of the so called $*$ -- eigenvalue equation. However this integral equation is in general difficult to be solved and, as with other physical theories, certain approximation methods are needed to obtain a solution.

Our intention is to propose an adaptation of the WKB method in order to obtain approximate solutions of a $*$ -- eigenvalue equation for a Hamilton function in  the phase space description of quantum physics. In this approach, due to the fact that the positions and momenta coordinates are used on an equal footing, a straightforward recalculating of the quasi -- classical formulas fails and  some additional  considerations are required. For example, the Weyl correspondence connecting wave functions and their respective Wigner functions is nonlocal.  Thus  we are faced to the problems of representing a product and a sum of wave functions in terms of the respective Wigner functions. Solutions of these two problems are analysed in this article. We also focus attention on an important question of representing an interference of wave functions as a contribution to the Wigner function.

The considerations presented in this work can be related to the problem of a semiclassical limit of a Wigner function \cite{ber}--\cite{dit}. However, we do not have to handle with singularities of Wigner functions, because difficulties caused by the term $\frac{1}{\hbar}$ are eliminated at the level of wave functions. It is worth to mention that in another context an application of the WKB expansion in formal deformation quantization has been studied in \cite{bord1,bord2}.

Our paper is organised as follows. In Sec. 2 the energy eigenvalue problem in the deformation quantization formalism is briefly described as well as the most important aspects of this quantization approach. Then, in Sec. 3 we present a review of the main elements of the WKB construction. Next, in Sec. 4, we transform the quasi--classical approximation for wave functions into an approximation for the respective Wigner functions. Specifically, we propose formulas to represent a product of wave functions and a sum of wave functions with separate supports. We introduce a Wigner function representing  interference of states and analyse some of its properties. Two examples of this WKB construction are presented in Sec. 5. General considerations are illustrated by calculations done for a one dimensional (1--D) harmonic oscillator and for unbound states in the Poeschl -- Teller potential. There are also two Appendices. The first one  contains a derivation of the Wigner function for states represented by tempered distributions. In the second appendix the Wigner functions of unbound states in the Poeschl -- Teller potential have been calculated. Finally, we give our concluding remarks.

\section{The energy eigenstates problem in deformation quantization}

The phase space quantum description is an alternative approach to the Hilbert space formalism of quantum mechanics. In this formalism the corresponding  space of states is a symplectic manifold and observables are represented by smooth real functions. Thus
 the coordinates and momenta are considered simultaneously, which gives a natural extension of the Hamiltonian construction to describe quantum systems.
 The  idea  is based on the fundamental observation that quantum physics is a deformed version of classical theory (see \cite{Bayen:1977ha}) where the role of deformation parameter is played by the Planck constant $\hbar$.

A general way to obtain the phase space quantum description of a system is through the deformation quantization formalism. Its main advantage is that it can be used to treat systems with arbitrary phase spaces although, since the multiplication of functions is replaced by a new product called the $*$ -- product, equations appearing in the formalism are in general difficult to  solve.

Nevertheless, the deformation formulation of quantum mechanics liberates us from numerous formal obstacles. First of all we do not need to construct an associated  Hilbert space of the system. We also avoid  quantization of observables and the nontrivial problem of defining domains of the constructed operators.
In its general version the deformation quantization calculus works on nontrivial symplectic spaces whereas other quantization procedures may not be established at all. Thus e.g. the $*$ -- eigenvalue equation (\ref{9}) and (\ref{9.1}) presented below is well defined
on an arbitrary symplectic manifold while the stationary Schroedinger being its counterpart may be not known.

 Our goal  is to obtain  an approximated method to find eigenvalues and eigenstates of a Hamilton function in the framework of deformation quantization. Thus we start from  the  $*$ -- eigenvalue energy employed in this approach.

The $*$ -- eigenvalue equation for a Hamilton function $H(\vec{r},\vec{p}\,)$ has  the following form
 \setcounter{orange}{1}
\renewcommand{\theequation} {\arabic{section}.\arabic{equation}\theorange}
\be
\label{9}
H(\vec{r},\vec{p}\,) * W_E(\vec{r},\vec{p}\,)= E \,W_E(\vec{r},\vec{p}\,),
\ee
where $E$ denotes an energy eigenvalue and  $W_E(\vec{r},\vec{p}\,)$ the corresponding Wigner energy eigenfunction.
Moreover, the additional condition
\addtocounter{orange}{1}
\addtocounter{equation}{-1}
\be
\label{9.1}
 \{H(\vec{r},\vec{p}\,), W_E(\vec{r},\vec{p}\,)\}_{\rm M}=0
 \ee
 \renewcommand{\theequation} {\arabic{section}.\arabic{equation}}
on the Wigner energy eigenfunction $W_E(\vec{r},\vec{p}\,)$ is imposed \cite{ja2}.

On the phase space ${\mathbb R}^{6},$ as the $*$ -- product we use the {\bf Moyal product} \cite{GW46,pleban,ja1}
\[
A(\vec{r},\vec{p}\,)* B(\vec{r},\vec{p}\,):=
\frac{1}{(\pi \hbar)^6 } \int_{{\mathbb R}^{12}} d\vec{r}\,'d\vec{p}\,'d\vec{r}\,'' d\vec{p}\,'' A(\vec{r}\,',\vec{p}\,') B(\vec{r}\,'',\vec{p}\,'')
\]
\be
\label{9.2}
\times \exp \left[ \frac{2i}{\hbar}\Big\{
(\vec{r}\,''-\vec{r}\,)\cdot(\vec{p}\,'-\vec{p}\,) -(\vec{r}\,'-\vec{r}\,)\cdot(\vec{p}\,''-\vec{p}\,)\Big\}
\right],
\ee
where
the dot `$\cdot$' stands for the scalar product. The above definition of the $*$ -- product is valid for a wide class of tempered distributions (for details see \cite{gracia}). This observation  is important since the Wigner energy eigenfunction $W_E(\vec{r},\vec{p}\,)$ can be a generalised function.

The sign convention employed in this paper is compatible with the Fedosov works \cite{6,7}.

The Moyal product is associative but in general non--Abelian. Moreover, it is closed i.e.
 \[
 \int_{{\mathbb R}^6} A(\vec{r},\vec{p}\,)*B(\vec{r},\vec{p}\,)d\vec{r}\,d\vec{p}=
 \int_{{\mathbb R}^6} B(\vec{r},\vec{p}\,)*A(\vec{r},\vec{p}\,)d\vec{r}\,d\vec{p}=
 \int_{{\mathbb R}^6} A(\vec{r},\vec{p}\,)\cdot B(\vec{r},\vec{p}\,)d\vec{r}\,d\vec{p}.
 \]
The {\bf Moyal bracket} appearing in the condition (\ref{9.1}) is defined as
\be
\label{11}
 \{A(\vec{r},\vec{p}\,), B(\vec{r},\vec{p}\,)\}_{\rm M}:= \frac{1}{i \hbar}\Big(A(\vec{r},\vec{p}\,)*B(\vec{r},\vec{p}\,) - B(\vec{r},\vec{p}\,)*A(\vec{r},\vec{p}\,)\Big).
\ee
From these definitions it can be deduced that formulas (\ref{9}) and (\ref{9.1}) are integral equations and  there is no universal way of solving them. Thus our idea is to  adapt the WKB approximation from the Hilbert space formulation of quantum mechanics to obtain approximated Wigner energy eigenfunctions. Such adaptation seems to be possible because for systems with phase spaces of the type ${\mathbb R}^{2n}$ a correspondence between the phase space description and the Hilbert space description is known.

Indeed, if we restrict to the  case of the phase space ${\mathbb R}^2,$ the
  phase space counterpart $A(x,p)$ of an operator $\hat{A}$ acting in the Hilbert space $L^2({\mathbb R})$
equals
\setcounter{orange}{1}
\renewcommand{\theequation} {\arabic{section}.\arabic{equation}\theorange}
\be
\label{12.1}
A(x,p)=
{\bf W}^{-1}\Big( \hat{A}\Big)=
 \int^{+\infty}_{-\infty} d \xi \,
 \Big< x - \frac{\xi}{2} \Big| \hat{A}
\Big| x + \frac{\xi}{2} \Big>  \, \mbox{exp} \left( -\frac{i\xi p}{\hbar}
\right).
\ee
\addtocounter{orange}{1}
\addtocounter{equation}{-1}
A similar formula in the momentum representation holds
\be
\label{12.2}
A(x,p)=
{\bf W}^{-1}\Big( \hat{A}\Big)=
 \int^{+\infty}_{-\infty} d \eta \,
 \Big< p - \frac{\eta}{2} \Big| \hat{A}
\Big| p + \frac{\eta}{2} \Big>  \, \mbox{exp} \left( \frac{i\eta x}{\hbar}
\right).
\ee
\renewcommand{\theequation} {\arabic{section}.\arabic{equation}}
The mapping ${\bf W}^{-1}$ is called the Weyl correspondence (for details see \cite{ja1}).
For every operator $\hat{A}$ such that its representation $\Big< x - \frac{\xi}{2} \Big| \hat{A}
\Big| x + \frac{\xi}{2} \Big>$ is a tempered generalised function of $\xi,$ the generalised function $A(x,p)$ is well defined. Note that the Weyl correspondence (\ref{12.1}, \ref{12.2}) is nonlocal.

 Applying the Weyl correspondence ${\bf W}^{-1}$  to a density operator of an energy eigenstate $|\psi_E\big> $ with the wave function $ \psi_{E}(x)$ it can be observed that the respective Wigner function is of the form
\be
\label{12}
W_E(x,p):=
{\bf W}^{-1}\Big( \frac{1}{2 \pi \hbar} |\psi_E\big>\big<\psi_E| \Big)=
\frac{1}{2 \pi \hbar} \int^{+\infty}_{-\infty} d \xi \,
\overline{\psi}_E \left( x + \frac{\xi}{2} \right)
\psi_E \left( x - \frac{\xi}{2} \right)  \mbox{exp} \left(- \frac{i\xi p}{\hbar}
\right).
\ee
Again the relation between the Wigner function and its respective wave function is nonlocal. The value of $W_E(x,p)$ at a fixed point $(x,p)$ depends on values of the function $\psi_{E}(x)$ on the whole configuration space. A strict form of expression (\ref{12}) applied to tempered generalised functions can be found in Appendix A.

\section{The WKB approximation in wave quantum mechanics \cite{daw,lan} }
\label{sec3}
\setcounter{equation}{0}

Let us consider the one particle nonrelativistic Schroedinger
equation
\be
\label{0}
- \frac{\hbar^2}{2M} \Delta \psi(t, \vec{r}\,)+ V(\vec{r}\,) \psi(t,\vec{r}\,) = i \hbar \frac{\partial \psi(t,\vec{r}\,)}{\partial t}.
\ee
If the potential $V(\vec{r}\,)$ does not depend on time, the stationary wave function $\psi(t,\vec{r}\,)$ is of the form
\be
\label{0.1}
\psi(t,\vec{r})=  \exp \left(- \frac{iEt}{\hbar}  \right) \psi_E(\vec{r}\,),
\ee
where $E$ denotes an eigenvalue of the Hamilton operator and $\psi_E(\vec{r}\,)$  its corresponding eigenfunction.
Hence $\psi_E(\vec{r}\,)$ obeys the stationary Schroedinger equation
\be
\label{0.2}
- \frac{\hbar^2}{2M} \Delta \psi_E( \vec{r}\,)+ V(\vec{r}\,) \psi_E(\vec{r}\,) = E \psi_E(\vec{r}\,).
\ee
Moreover,
every solution of (\ref{0.2}) can be written as a linear combination of two functions
\be
\label{2}
 \psi_{E\, I} (\vec{r}\,)=\exp \left( \frac{i}{\hbar} \sigma_{I}(\vec{r}\,) \right) \;\;\; {\rm and } \;\;\; \psi_{E\, II} (\vec{r}\,)=\exp \left( \frac{i}{\hbar} \sigma_{II}(\vec{r}\,) \right)
\ee
satisfying separately Eq. (\ref{0.2}),
where $\sigma_{I}(\vec{r}\,)$ and $ \sigma_{II}(\vec{r}\,)$ denote some complex valued functions.
The phases $\sigma_{I}(\vec{r}\,)$ and $\sigma_{II}(\vec{r}\,)$ fulfill the  second order nonlinear partial   differential equation
\be
\label{3}
\frac{1}{2M}\big(\nabla \sigma(\vec{r}\,)\big)^2 - \frac{i \hbar}{2M} \Delta \sigma(\vec{r}\,)= E - V(\vec{r}\,) ,
\ee
for $ \sigma(\vec{r}\,) = \sigma_{I}(\vec{r}\,) $ and $ \sigma(\vec{r}\,) = \sigma_{II}(\vec{r}\,).$
In the classical limit $\hbar \rightarrow 0$ this equation reduces to the Hamilton -- Jacobi stationary equation
\be
\label{4}
\frac{1}{2M}\big(\nabla \sigma(\vec{r}\,)\big)^2 = E - V(\vec{r}\,),
\ee
where the function  $\sigma(\vec{r}\,)$ is interpreted as the stationary action and its partial derivatives $\frac{\partial \sigma(\vec{r}\,)}{ \partial x}, \frac{\partial \sigma(\vec{r}\,)}{\partial y}, \frac{\partial \sigma(\vec{r}\,)}{\partial z} $ are the momenta.

The expression (\ref{3}) is equivalent to the stationary Schroedinger equation (\ref{0.2}). The phases $ \sigma_{I}(\vec{r}\,)\, , \, \sigma_{II}(\vec{r}\,)$ carry the same information about the eigenstate of the Hamilton operator  as the energy eigenfunction does.  However, it is usually more difficult to solve (\ref{3}) than  Eq. (\ref{0.2}). Importance of the formula (\ref{3}) lies in an iterative procedure in order to derive the function    $ \sigma(\vec{r}\,).$

For simplicity let us consider the $1$--D case. Then Eq. (\ref{3}) is of the form
\be
\label{5}
\frac{1}{2M}\left(\frac{ d\sigma(x)}{dx}\right)^2 - \frac{i \hbar}{2M} \frac{ d^2\sigma(x)}{dx^2}= E - V(x).
\ee
In certain parts of its domain
the solution can be written as a formal power series in the Planck constant
\be
\label{6}
\sigma(x)= \sum_{k=0}^{\infty} \left( \frac{\hbar}{i}\right)^k \sigma_k(x).
\ee

Thus we receive an iterative system of equations
\bea
\label{7}
\frac{1}{2M} \left(\frac{ d\sigma_0(x)}{dx}\right)^2 & = & E-V(x), \nonumber  \vspace{0.25cm}\\
 \frac{ d\sigma_0(x)}{dx} \frac{ d\sigma_1(x)}{dx} + \frac{1}{2}\frac{ d^2\sigma_0(x)}{dx^2} & = & 0, \nonumber  \vspace{0.25cm} \\
\frac{ d\sigma_0(x)}{dx} \frac{ d\sigma_2(x)}{dx}+ \frac{1}{2}  \left(\frac{ d\sigma_1(x)}{dx}\right)^2 + \frac{1}{2} \frac{ d^2\sigma_1(x)}{dx^2}
& = & 0,   \vspace{0.25cm} \\
\vdots & \vdots & \vdots \nonumber
\eea
The  element $\sigma_0(x)$ is the classical stationary action for the system while the next terms can be interpreted as quantum corrections.

The recurrence (\ref{7}) leads to the following set of conditions
\bea
\label{7.0}
\frac{ d\sigma_0(x)}{dx} &=& \pm \sqrt{2M(E-V(x))}  \nonumber  \vspace{0.25cm}\\
\vdots & \vdots & \vdots \nonumber \\
\frac{ d\sigma_n(x)}{dx}&=& \frac{1}{\frac{ d\sigma_0(x)}{dx}} \cdot f_n \left( \frac{ d^2\sigma_{n-1}(x)}{dx^{2}} , \frac{ d\sigma_{n-1}(x)}{dx}, \ldots,  \frac{ d\sigma_1(x)}{dx}\right) \\
\vdots & \vdots & \vdots \nonumber
\eea
By $f_n(y_1, \ldots,y_n)$ we denote a polynomial in variables $y_1, \ldots, y_n.$

There exist two solutions of Eqs. (\ref{7.0}) and they differ on the sign at even $\hbar$ power elements. Thus the phases $\sigma_{I}(x)$ and
$\sigma_{II}(x)$ from the formula (\ref{2}) are
\be
\label{7.1}
\sigma_{I}(x)= \sum_{k=0}^{\infty} \left( \frac{\hbar}{i}\right)^k \sigma_k(x) \; , \;\;\;
\sigma_{II}(x)= \sum_{k=0}^{\infty} \left( \frac{\hbar}{i}\right)^k (-1)^{k+1}\sigma_k(x).
\ee
The odd coefficients $\sigma_{2k+1}$  can be selected real and  the even elements $\sigma_{2k}, \;k=0,1,2, \ldots$  can be chosen real for $E-V(x)>0$ and  imaginary for $E-V(x)<0.$ This freedom of choice results from the fact that the system (\ref{7}) determines only the first derivatives of $\sigma_{k}, k=0,1,2, \ldots$ Hence an arbitrary complex number can be added to each term $\sigma_{k}.$ Thus the  energy eigenfunction is normalisable and
 one can write
 \setcounter{orange}{1}
\renewcommand{\theequation} {\arabic{section}.\arabic{equation}\theorange}
\be
\label{7.2}
\sigma_{I}(x)=\sigma_{\rm odd}(x) + \sigma_{\rm even}(x) \; , \;\;\;
\sigma_{II}(x)=\sigma_{\rm odd}(x) - \sigma_{\rm even}(x),
\ee
\addtocounter{orange}{1}
\addtocounter{equation}{-1}
where
\be
\label{7.2a}
\sigma_{\rm odd}(x):= \sum_{k=0}^{\infty} \left( \frac{\hbar}{i}\right)^{2k+1} \sigma_{2k+1}(x) \; ,\;\;\;
\sigma_{\rm even}(x) := \sum_{k=0}^{\infty} \left( \frac{\hbar}{i}\right)^{2k} \sigma_{2k}(x).
\ee
\renewcommand{\theequation} {\arabic{section}.\arabic{equation}}

With respect to the choice: $\sigma_{2k+1}(x)$ -- real,  $\sigma_{2k}(x)$ -- real or  imaginary we conclude that $\sigma_{\rm odd}(x)$ is an imaginary function and $\sigma_{\rm even}(x) $ is a real function for $E-V(x)>0$ and an  imaginary function for $E-V(x)<0.$

The series expansion cannot be applied in neighbourhoods of turning points. The system  (\ref{7}) consists of infinitely many equations. For obvious purposes we would like to restrict ourselves to a finite number of terms in the series (\ref{6}). However, it is possible only if a few initial terms make a principal contribution to the sum (\ref{6}). Thus the necessary but not sufficient condition   for  the expansion (\ref{6}) to be  applicable is   that it should be taken at points $x$ satisfying  the inequality
\be
\label{8}
|x-x_0| \gg \frac{1}{2} \left( \frac{\hbar^2}{M \Big| \frac{dV(x)}{dx}\big|_{x=x_0}\Big|}\right)^{\frac{1}{3}}
\ee
(for details see \cite{daw,lan}). Certainly the  turning points never obey (\ref{8}) and their neighbourhoods must be analysed separately.

One has to remember about another restriction imposed on the expansion (\ref{6}).
 The turning points are places at which the momentum $p= \pm \sqrt{E-V(x)}$ is zero. Hence from the first equation of the system (\ref{7}) we see that the expansion (\ref{6}) may not   work at  points at which the classical momentum is small.

In order to find a complete approximate solution of the stationary Schroedinger equation we have to
solve first the system of equations (\ref{7})  in all intervals, in which the series expansion (\ref{6}) is acceptable. Then we
 match these  approximate solutions.  There are several techniques of making the separate approximate solutions compatible \cite{fro, gal}. The one used here applies strict solutions of the stationary Schroedinger equation near turning points.

The algorithm sketched above is widely used and appears in the literature as the {\it quasi -- classical approximation} or the {\it Wentzel -- Kramers -- Brillouin (WKB)  approximation}.

\begin{figure}[h]\centering
\includegraphics[scale=.5]{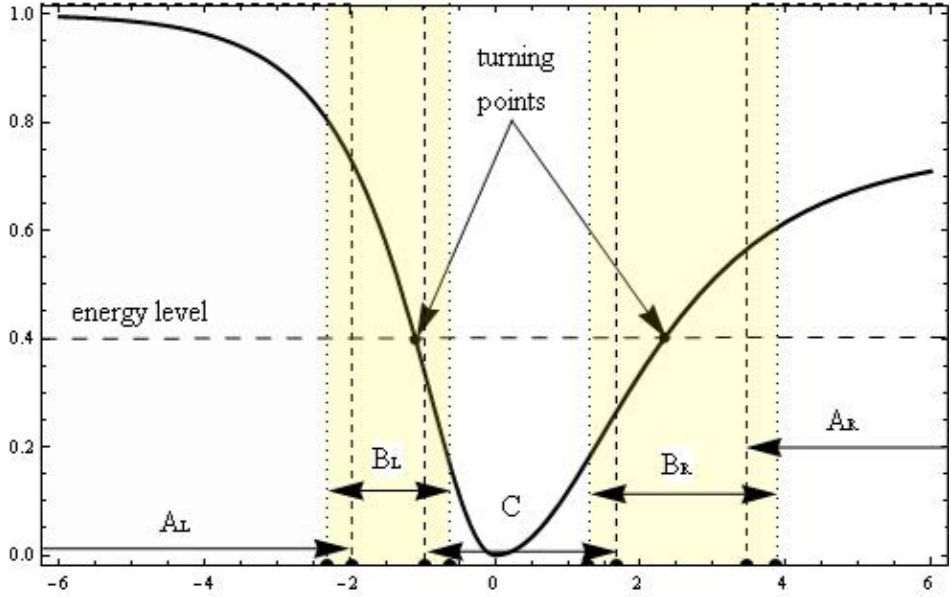}
\caption{A potential $V(x)$ as a function of $x$. }
\label{pic1.00}
\end{figure}

Let us  discuss its application to a $1$--D potential  drawn on    FIG. \ref{pic1.00}. We look for a solution of the Schroedinger stationary equation  using the first order of the quasi--classical approximation.

The domain of Eq. (\ref{5}) has been divided in five regions belonging to three classes.
\begin{enumerate}
\item
Classically forbidden regions $A_L=(- \infty,a_{Lr}]$ and $A_R=[a_{Rl}, \infty)$. In these regions the iterative system of equations (\ref{7}) is applicable.
\item
Neighbourhoods $B_L=[b_{Ll}, b_{Lr}]$  and $B_R=[b_{Rl},b_{Rr}]$ of the turning points in which is necessary to approximate the potential $V(x)$ by a polynomial and then to find a strict solution of a $1$--D version of the stationary Schroedinger equation (\ref{0.2}).
\item
A part $C=[c_l,c_r]$  of the classically accessible area between the turning points $x_1$ and $x_2,$ in which  the iterative procedure (\ref{7}) can be used.
\end{enumerate}
In the area $A_L$  the energy $E$ is smaller than the potential $V(x).$  From the first equation of the system (\ref{7}) we obtain that
\be
\label{22}
\sigma_{0}(x)= \pm i \int_{x_1}^x {\bf p}(y)dy\;, \;\; {\bf p}(y):= \sqrt{2M(V(y)-E)}.
\ee
 The choice of the sign results from the requirement that the  spatial probability of detection of the particle  must be normalisable. Thus the unique physically acceptable first order solution of the system (\ref{7}) is the pair of functions
\be
\label{23}
\sigma_{0}(x)= - i \int_{x_1}^x {\bf p}(y)dy \;,\;\; \;\sigma_1(x)= \ln \frac{1}{\sqrt{{\bf p}(x)}}+G_1, \; G_1 \in {\mathbb C}.
\ee
The respective wave function equals
\be
\label{23.1}
\psi_{E(1)A_L}(x)= \frac{D_{(1)A_L}}{\sqrt{{\bf p}(x)}}\exp \left( \frac{1}{\hbar} \int_{x_1}^x {\bf p}(y)dy \right),
\ee
where $D_{(1)A_L}$ corresponds to a normalising factor.

For the region $B_L$  the potential $V(x)$ is approximated by a polynomial.
 Since we consider only the first step of the quasi--classical approximation, it is sufficient to assume that  the potential is linear
\be
\label{27}
V(x)\approx V(x_1) - F_{x_1}(x-x_1)\;, \;\;\; F_{x_1} := - \frac{d V(x)}{dx} \Big|_{x=x_1}.
\ee
In this case the stationary Schroedinger equation reduces to the form
\be
\label{27.5}
\frac{d^2 \psi_{EB_L}(x)}{dx^2}+ \frac{2MF_{x_1}}{\hbar^2}(x-x_1) \psi_{E B_L}(x)=0.
\ee
The probability of detection the particle diminishes as the coordinate $x $ decreases. Thus the unique
  acceptable solution of (\ref{27.5}) is
\be
\label{28}
\psi_{E B_L}(x)= C_{B_L} \Phi \left( \left(\frac{2MF_{x_1}}{\hbar^2}\right)^{1/3}(x_1-x)\right),
\ee
where $\Phi(y)$ denotes the Airy function defined by  $\Phi(y):= \frac{1}{\sqrt{\pi}} \int_0^{\infty} \cos\left(\frac{u^3}{3}+ uy \right)du$ and the constant $ C_{B_L} \in {\mathbb C}.$

In the intersection area of regions $A_L$ and $B_L,$  where $x_1-x>>0$ and  the relationship (\ref{8}) holds, an asymptotic expansion of the solution (\ref{28}) can be applied (for details see \cite{lib}).
The asymptotic expansion of (\ref{28}) and the function (\ref{23.1}) coincide under the condition $D_{(1)A_L}= \frac{C_{B_L}}{2}(2 M F_{x_1}\hbar)^{1/6}. $

The region $C$ is classically accessible and the wave function in this area can be chosen real so in the first quasi--classical approximation we have
\be
\label{32}
\psi_{E ( 1) C}(x)= \frac{D_{(1) C}}{\sqrt{{\bf p}(x)}}\sin \left( \frac{1}{\hbar}\int_{x_1}^x {\bf p}(y)dy + \delta \right), \; {\bf p}(y):= \sqrt{2M(E-V(y))},\; D_{(1) C}, \delta \in {\mathbb R}.
\ee
Inside the intersection of the regions $C$ and $B_L,$ where $x_1 - x <<0$ and the condition (\ref{8}) is fulfilled it can be employed the asymptotic expansion of (\ref{28}).

Therefore we can see that
\be
\label{36}
\delta= \frac{\pi}{4} + k \pi \;, \;\;\; k \in {\mathbb Z} \;\;\;\mbox{and} \;\;\; D_{(1) C}= (-1)^k C_{B_L} (2 M F_{x_1}\hbar)^{1/6}.
\ee
If we take the simplest choice $k=0,$ the wave function is given by
\be
\label{38}
\psi_{E (1) C}(x)= \frac{D_{(1) C}}{\sqrt{{\bf p}(x)}}\sin \left( \frac{1}{\hbar}\int_{x_1}^x {\bf p}(y)dy + \frac{\pi}{4} \right).
\ee
In a  way analogous to the earlier considerations we obtain that  in the area $A_R$
 the wave function equals
\be
\label{38.2}
\psi_{E(1)A_R}(x)= \frac{D_{(1)A_R}}{\sqrt{{\bf p}(x)}}\exp \left( -\frac{1}{\hbar} \int_{x_2}^x {\bf p}(y)dy \right).
\ee

In a neighbourhood of the turning point $x_2$ we have a wave function  similar to (\ref{28})
\be
\label{40}
\psi_{E B_R}(x)= C_{B_R} \Phi \left( \left(\frac{2MF_{x_2}}{\hbar^2}\right)^{1/3}(x_2-x)\right)\;, \;\;\;
F_{x_2} := - \frac{d V(x)}{dx} \Big|_{x=x_2}.
\ee
The coincidence condition between the solutions in the areas $A_R$ and $B_R$ means that there must be $D_{(1)A_R}= \frac{C_{B_R}}{2}(-2 M F_{x_2}\hbar)^{1/6}.$

Finally the wave function in the region $C$ seen from the side of the turning point $x_2$ is
\be
\label{41}
\psi_{E ( 1) C}'(x)= \frac{D_{(1) C}'}{\sqrt{{\bf p}(x)}}\sin \left(- \frac{1}{\hbar}\int_{x_2}^x {\bf p}(y)dy + \delta' \right), \;{\bf p}(y):= \sqrt{2M(E-V(y))},\; D_{(1) C}', \delta'\in {\mathbb R}.
\ee
As before,  comparing the approximate solutions in the intersection $C \cap B_R$ we conclude that
\be
\label{42}
\delta'= \frac{\pi}{4} + k \pi\; , \;\;\; k \in {\mathbb Z} \;\;\; \mbox{and} \;\;\; D_{(1) C}'= (-1)^k C_{B_R} (-2 M F_{x_2}\hbar)^{1/6}.
\ee
Again, selecting $k=0$ we  obtain
\be
\label{44}
\psi_{E (1) C}'(x)= \frac{D_{(1) C}'}{\sqrt{{\bf p}(x)}}\sin \left( -\frac{1}{\hbar}\int_{x_2}^x {\bf p}(y)dy + \frac{\pi}{4} \right).
\ee
The functions (\ref{38}) and (\ref{44}) must be equal.
Since the integral $\int_{x_1}^{x_2}{\bf p}(y)dy $ is positive, we arrive to the well known quantization rule
\be
\label{45}
\int_{x_1}^{x_2}{\bf p}(y)dy = \hbar \left( n + \frac{1}{2} \right) \pi\; , \;\;\; n \in {\cal N}.
\ee
The coefficients $D_{(1) C}'$ and $D_{(1) C}$ fulfill the equality $D_{(1) C}= (-1)^n D_{(1) C}'.$

In the next chapter the elements of the WKB construction presented here will be extended to the deformation quantization formalism.


\section{An adaptation of the WKB approximation to deformation quantization}

\setcounter{equation}{0}

The main idea of an approximation method in the WKB approach to solve  the energy eigenvalue problem (\ref{9}, \ref{9.1}) for a Hamilton function $H(\vec{r}, \vec{p})$ is to explore the equation (\ref{3}) for a phase $\sigma(\vec{r}).$ In the $1$--D case a relationship between an energy Wigner eigenfunction $W_E(x,p)$ and the phase $\sigma(x)$ follows from the Weyl correspondence (\ref{12})
\be
\label{12.10}
W_E(x,p)=  \frac{1}{2 \pi \hbar} \int^{+\infty}_{-\infty} d \xi \, \mbox{exp} \left( \frac{i}{\hbar} \left[ \sigma \left(x- \frac{\xi}{2} \right)
- \overline{\sigma} \left(x+ \frac{\xi}{2} \right) -\xi p
\right] \right).
\ee
 Notice that the relation between the Wigner function  and the phase is global i.e. the value of $W_E(x,p)$ at a fixed point depends on values of the phase $\sigma(x)$ on the whole configuration space.

 Unfortunately, as we mentioned in the previous section, it is usually necessary to restrict to an approximated solution of Eq. (\ref{5}). In particular,  the iterative WKB algorithm of solving this problem has been discussed. Thus to propose an effective method of dealing with the energy eigenstates problem in deformation quantization we have to transform elements of the WKB approximation for wave functions into procedures dedicated to Wigner functions.

 In fact four elements must be considered:  conditions imposed on eigenvalues of the Hamilton function, representation of the sum phases in terms of phase space functions related to these phase, a relationship between a superposition of wave functions and their Wigner functions and finally, the connection formulas for partial Wigner functions.

 As it could be seen  in Sec. \ref{sec3}, there is no universal rule determining energy eigenvalues in the WKB method. For example in the first approximation for a potential like the one depicted at FIG. \ref{pic1.00} the  condition imposed on admissible values of energy is of the form (\ref{45}).

 Three other components of the WKB procedure require more detailed analysis. Thus to each of them a separate subsection will be devoted.

\subsection{The phase space representation of a product of wave functions}

Let us go back to the formula (\ref{6}), on which the WKB approximation is based. We see that in the case when the phase $\sigma(x)$ is a power series in the Planck constant, the wave function has the following form
\be
\label{13}
\psi_E(x)= \prod_{k=0}^{\infty} \psi_{E \,k}(x)\;\;, \;\; \psi_{E \,k}(x):= \exp \left[ \frac{i}{\hbar} \left(\frac{\hbar}{i} \right)^k \sigma_k(x) \right].
\ee
The functions $\psi_{E \,k}(x), \;k=0,1, \ldots $ need not be  elements of $L^2({\mathbb R})$ but as they are smooth and, due to physical requirements,  bounded, the product
$\overline{\psi}_{E \,k}\left( x + \frac{\xi}{2} \right)
\psi_{E \,k} \left( x - \frac{\xi}{2} \right)$ is a tempered generalised function for every $k$. Since the Weyl correspondence (\ref{12}) is in general a Fourier transform of the product of functions  $\overline{\psi} \left( x + \frac{\xi}{2} \right)
\psi \left( x - \frac{\xi}{2} \right)$ (look the Appendix \ref{appA}),
 then the integral $\int^{+\infty}_{-\infty} d \xi \,
\overline{\psi}_{E\,k} \left( x + \frac{\xi}{2} \right)
\psi_{E\,k} \left( x - \frac{\xi}{2} \right)  \mbox{exp} \left(- \frac{i\xi p}{\hbar}
\right)$ is well defined.

Under the conditions discussed in Sec. \ref{sec3} the finite product $ \prod_{k=0}^{n} \psi_{E \,k}(x)$ may be a good approximation of an energy eigenstate.
This approximation can be realised as an iterative procedure, in which the $n$-th approximation $\psi_{E(n)}(x)$ of the wave function $\psi_E(x)$ equals
\be
\label{14}
\psi_{E(0)}(x):= \psi_{E0}(x) \;\;\;, \;\;\;
\psi_{E(n)}(x)= \psi_{E(n-1)}(x)\cdot \psi_{E\,n}(x)\;, \;n \geq 1.
\ee
Thus we are interested in finding a formula expressing $W_{E(n)}(x,p):={\bf W}^{-1}\Big( \frac{1}{2 \pi \hbar} |\psi_{E(n)}\big>\big<\psi_{E(n)}| \Big)$ by
$W_{E(n-1)}(x,p):={\bf W}^{-1}\Big( \frac{1}{2 \pi \hbar} |\psi_{E(n-1)}\big>\big<\psi_{E(n-1)}| \Big)$ and
$W_{E\,n}(x,p):={\bf W}^{-1}\Big( \frac{1}{2 \pi \hbar} |\psi_{E \,n }\big>\big<\psi_{E \,n}| \Big).$
From (\ref{12}) it can be observed that
\[
\int_{-\infty}^{+\infty}dp \, W_E(x,p) \mbox{exp} \left( \frac{i\lambda p}{\hbar}
\right)= \overline{\psi}_E \left( x + \frac{\lambda}{2} \right)
\psi_E \left( x - \frac{\lambda}{2} \right).
\]
This result in fact is true for an arbitrary pure state.
Now from (\ref{12}) and (\ref{14}) we obtain
\[
W_{E(n)}(x,p)= \frac{1}{2 \pi \hbar} \int_{-\infty}^{+\infty}dp' \int_{-\infty}^{+\infty}dp'' \int_{-\infty}^{+\infty}d\lambda \, W_{E(n-1)}(x,p') W_{E\,n}(x,p'') \mbox{exp} \left( \frac{i\lambda (p'+p''-p)}{\hbar} \right)
\]
\be
\label{15}
=  \int_{-\infty}^{+\infty} dp' W_{E(n-1)}(x,p') W_{E\,n}(x,p-p')=  \int_{-\infty}^{+\infty} dp'' W_{E(n-1)}(x,p-p'') W_{E\,n}(x,p'').
\ee
In this way the  Wigner function corresponding to the product of two arbitrary wave functions  is represented by the convolution of their Wigner functions with respect to the momentum.

Analogously, if we work in the momentum representation with a function of the form
\be
\label{15.1}
\psi_{E(0)}(p):= \psi_{E0}(p) \;, \;\;\;
\psi_{E(n)}(p)= \psi_{E(n-1)}(p)\cdot \psi_{E\,n}(p)\;, \;n \geq 1,
\ee
then
\be
\label{15.2}
W_{E(n)}(x,p)= \int_{-\infty}^{+\infty} dx' W_{E(n-1)}(x',p) W_{E\,n}(x-x',p)
 =  \int_{-\infty}^{+\infty} dx'' W_{E(n-1)}(x-x'',p) W_{E\,n}(x'',p).
\ee
\subsection{The phase space counterpart of a superposition of wave functions}

The main difficulty in the application of the WKB method is that
the series expansion (\ref{6}) cannot be applied everywhere. Thus in the quasi--classical approximation the total wave function is represented by a sum of  $k$ spatially separable functions
\be
\label{15.3}
\psi_E(x)= \sum_{l=1}^k \psi_{E a_l b_l}(x) \;\;\; ,\;\;\; - \infty \leq a_1 < b_1=a_2 < b_2=a_3 < \ldots <b_{k-1}= a_k < b_k \leq \infty.
\ee
The function $\psi_E(x)$ need not be continuous at the points $a_i, b_i, \; i=1,2, \ldots k.$ Nevertheless,
every partial function $\psi_{E a_l b_l}(x)$ satisfies the condition ${\rm supp} \, \psi_{E a_l b_l}(x) \subseteq [a_l,b_l].$ Therefore
\[
\psi_{E a_l b_l}(x)= Y(x-a_l) \psi_{E a_l b_l}(x) Y(b_l-x)=Y(x-a_l) \psi_E(x) Y(b_l -x),
\]
where $Y(x)$ denotes the Heaviside function. In this way the set of functions $\{\psi_{E a_l b_l}(x)\}_{l=1}^{k}$ is orthogonal. Provided the global function
$\psi_E(x)$ is normalised, we obtain
$
\sum_{l=1}^k ||\psi_{E a_l b_l}(x)||^2=1.
$

Defining the self--adjoint operators
\be
\label{15.30}
\widehat{Pr}_{E a_l b_l}:= \frac{1}{\big<\psi_{E a_l b_l}|\psi_{E a_l b_l} \big>} |\psi_{E a_l b_l} \big> \big< \psi_{E a_l b_l}|, \; l=1, \ldots, k
\ee
which are projectors on $1$--D closed subspaces of a Hilbert space ${\cal H},$ one can see that their product satisfies the relation
\[
\widehat{Pr}_{E a_l b_l} \widehat{Pr}_{E a_r b_r} = \delta_{lr} \widehat{Pr}_{E a_r b_r}.
\]
It is now of  vital importance to obtain  a phase space counterpart of the state being the superposition of wave functions of the form (\ref{15.3}). Let us consider a Wigner function arising from a wave function $\psi_{E a_l b_l}(x).$  Applying (\ref{12}) one gets
\be
\label{16}
W_{E a_l b_l}(x,p)= \frac{1}{2 \pi \hbar}\int_{{\rm Max.}[2(a_l-x),2(x-b_l)]}^{{\rm Min.}[2(x-a_l),2(b_l-x)]} d \xi\,
\overline{\psi}_{E a_l b_l} \left( x + \frac{\xi}{2} \right)
\psi_{E a_l b_l} \left( x - \frac{\xi}{2} \right)  \mbox{exp} \left( -\frac{i\xi p}{\hbar}
\right).
\ee
Therefore
the Wigner function $W_{E a_l b_l}(x,p)$ vanishes  outside the set $(a_l,b_l) \times {\mathbb R}.$ As the function $\psi_{E a_l b_l}(x)$ itself can be a sum of functions, we observe that every  Wigner function corresponding to a superposition of wave functions
with supports from an interval $[a_l, b_l]$
is still limited to the strip $a_l \leq  x  \leq b_l.$

Moreover,
for $a_l \leq x \leq  \frac{a_l+b_l}{2}$ the integral (\ref{16}) turns into
\[
\int_{2(a_l-x)}^{2(x-a_l)} d \xi\,
\overline{\psi}_{E a_l b_l} \left( x + \frac{\xi}{2} \right)
\psi_{E a_l b_l} \left( x - \frac{\xi}{2} \right)  \mbox{exp} \left(- \frac{i\xi p}{\hbar}
\right)
\]
and for $ \frac{a_l+b_l}{2} \leq  x \leq  b_l$ we obtain
\[
\int_{2(x-b_l)}^{2(b_l-x)} d \xi\,
\overline{\psi}_{E a_l b_l} \left( x + \frac{\xi}{2} \right)
\psi_{E a_l b_l} \left( x - \frac{\xi}{2} \right)  \mbox{exp} \left( -\frac{i\xi p}{\hbar}
\right).
\]
The interval of integration always contains $0,$ is symmetric with respect to the point $\xi=0$ and its length increases from $0$ for $x=a_l$ to $2(b_l-a_l)$ for $x=\frac{a_l+b_l}{2}.$ Then it decreases to $0$ for $x=b_l.$  As a Wigner function, the function $W_{E a_l b_l}(x,p)$ is continuous with respect to $x$ and $p$ and finite at every point of ${\mathbb R}^2.$ If $a_l$ and $b_l$ are finite, the function  $W_{E a_l b_l}(x,p)$ as  the Fourier transform of a function with compact support, is smooth and its support is unbounded.

Analogously, for every wave function of the form $\psi_{E c_l d_l}(p)= Y(p-c_l)\psi_E(p)Y(d_l-p), \; c_l < d_l$ in the momentum representation its Wigner function $W_{E c_l d_l}(x,p)$ vanishes for $p \leq c_l$ and $p \geq d_l.$

From the expression (\ref{12.1}) we can deduce that if an operator $\hat{A}$   in the position representation satisfies the condition
$
\big<x| \hat{A}|x'\big> \neq 0 \;\;\; {\rm only \;\; for} \;\;\; a < x,x' < b,
$
then  the function ${\bf W}^{-1}(\hat{A})(x,p)$ may be different from $0$  only for $x$  contained in the interval $(a,b).$ Moreover, the function ${\bf W}^{-1}(\hat{A})(x,p)$ is a smooth function  with respect to the momentum $p.$ For every $\tilde{x} \in (a,b) $  and every positive number $\Lambda >0$ there exists a value of momentum $\tilde{p}$ such that $|\tilde{p}|>\Lambda$ and  ${\bf W}^{-1}(\hat{A})(\tilde{x},\tilde{p}) \neq 0.$

An analogous conclusion can be formulated for every  operator $\hat{A}$ such that in the momentum representation
$
\big<p| \hat{A}|p'\big> \neq 0 \;\;\; {\rm only \;\; for} \;\;\; c < p,p' < d.
$

It is important to point out that the inverse statement is not true i.e. the fact that a function $A(x,p)= {\bf W}^{-1}(\hat{A})$ differs from $0$ only for $a < x < b $
or, respectively  $c < p < d ,$
does not imply that $\big<x| \hat{A}|x'\big> \neq 0 $ exclusively  for $ a < x,x' < b$ or, respectively $\big<p| \hat{A}|p'\big> \neq 0 $ exclusively for $c < p,p' < d .$

Indeed, let $A(x,p)= Y(x+a)Y(a-x)Y(p+c)Y(c-p)$ with $a,c>0.$ Then
\[
\big< x|\hat{A}|x' \big>= 2 \hbar \frac{\sin \left( \frac{c(x-x')}{\hbar}\right)}{x-x'}
\]
for $-2a < x+x' < 2a$ and $0$ for $x,x'$ not fulfilling these inequalities.

Finally, for every $1 \leq l, r \leq k$ one finds that
\[
\int_{a_l}^{b_l} dx \int_{-\infty}^{\infty} dp W_{E a_l b_l}(x,p)= || \psi_{E a_l b_l}||^2 \leq 1
\]
and
\[
\int_{a_l}^{b_l} dx \int_{-\infty}^{\infty} dp W_{E a_l b_l}(x,p) W_{E a_r b_r}(x,p) = \delta_{lr}\frac{|| \psi_{E a_l b_l}||^3}{2 \pi \hbar}.
\]

Now we are going to analyse a problem of representing a superposition of wave functions on the phase space. Let us consider a two-component linear combination of functions
\[
 Y(x-a_l)\psi_{Ea_l b_l}(x)Y(b_l-x) + Y(x-a_r)\psi_{E a_r b_r}(x)Y(b_r-x), \;\;\;\; - \infty \leq a_l < b_l \leq a_r < b_r \leq \infty.
\]
Its Wigner function is then given by
 \[
 W_E(x,p)= {\bf W}^{-1}\Big( \frac{1}{2 \pi \hbar} |\psi_{Ea_l b_l}\big>\big<\psi_{E a_l b_l}| \Big)
 +{\bf W}^{-1}\Big( \frac{1}{2 \pi \hbar} |\psi_{E a_r b_r}\big>\big<\psi_{E a_r b_r}| \Big)+
  \]
 \be
 \label{16.1}
 +{\bf W}^{-1}\Big( \frac{1}{2 \pi \hbar} |\psi_{E a_l b_l}\big>\big<\psi_{E a_r b_r}| + \frac{1}{2 \pi \hbar} |\psi_{E a_r b_r}\big>\big<\psi_{E a_l b_l}|\Big).
 \ee
 The components ${\bf W}^{-1}\Big( \frac{1}{2 \pi \hbar} |\psi_{Ea_l b_l}\big>\big<\psi_{E a_l b_l}| \Big)$ and
 ${\bf W}^{-1}\Big( \frac{1}{2 \pi \hbar} |\psi_{E a_r b_r}\big>\big<\psi_{E a_r b_r}| \Big)$ belong to the set of partial Wigner functions analysed before. However the other two terms are essentially new and they correspond to the function
\be
\label{17}
 W_{E \,int}(x,p):={\bf W}^{-1}\Big( \frac{1}{2 \pi \hbar} |\psi_{E a_l b_l}\big>\big<\psi_{E a_r b_r}| + \frac{1}{2 \pi \hbar} |\psi_{E a_r b_r}\big>\big<\psi_{E a_l b_l}|\Big),
\ee
which represents the interference between the partial wave functions $\psi_{Ea_l b_l}(x)$ and $\psi_{Ea_r b_r}(x).$

The operator appearing in (\ref{17}) and defined as
\be
\widehat{\rm Int}:=|\psi_{E a_l b_l}\big>\big<\psi_{E a_r b_r}| + |\psi_{E a_r b_r}\big>\big<\psi_{E a_l b_l}|
\ee
is self-adjoint although it is not a projector. Its trace vanishes and it has three possible eigenvalues $\lambda$:
   \setcounter{orange}{1}
\renewcommand{\theequation} {\arabic{section}.\arabic{equation}\theorange}
\be
\lambda_{-}=-||\psi_{E a_l b_l}|| \cdot ||\psi_{E a_r b_r}||\;\;, \;\;|-\big> = \frac{1}{\sqrt{2}}
\left( \frac{1}{||\psi_{E a_l b_l}|| } |\psi_{E a_l b_l}\big> - \frac{1}{ ||\psi_{E a_r b_r}||}|\psi_{E a_r b_r}\big>\right),
\ee
\addtocounter{orange}{1}
\addtocounter{equation}{-1}
\be
\lambda_{0}=0 \;\; , \;\; {\rm its \; eigenvector \; is \; every \; vector \; orthogonal\; to\; } |\psi_{E a_l b_l}\big> \; {\rm and}\; |\psi_{E a_r b_r}\big>,
\ee
\addtocounter{orange}{1}
\addtocounter{equation}{-1}
\be
\lambda_{+}=||\psi_{E a_l b_l}|| \cdot ||\psi_{E a_r b_r}||\;\;, \;\;| +\big> = \frac{1}{\sqrt{2}}
\left( \frac{1}{||\psi_{E a_l b_l}|| } |\psi_{E a_l b_l}\big> + \frac{1}{ ||\psi_{E a_r b_r}||}|\psi_{E a_r b_r}\big>\right).
\ee
\renewcommand{\theequation} {\arabic{section}.\arabic{equation}}
The interference operator $\widehat{\rm Int}$ exchanges the directions between vectors $|\psi_{E a_l b_l}\big> \rightleftharpoons  |\psi_{E a_r b_r}\big>.$ Indeed,
\[
\widehat{\rm Int} |\psi_{E a_l b_l}\big>= ||\psi_{E a_l b_l}||^2\,|\psi_{E a_r b_r}\big>\;\;\;, \;\;\;
\widehat{\rm Int} |\psi_{E a_r b_r}\big>= ||\psi_{E a_r b_r}||^2\,|\psi_{E a_l b_l}\big>.
\]

The function $W_{E \,int}(x,p)$ defined by the expression (\ref{17}) and representing the interference term is determined by the real part of the integral
\be
\label{18}
W_{E \,int}(x,p)=
 2\, {\rm Re}\, \left( \int_{{\rm Max.}[2(a_l-x),2(x-b_r)]}^{{\rm Min.}[2(b_l-x),2(x-a_r)]}
 d \xi \, \overline{\psi}_{E a_l b_l} \left(x+ \frac{\xi}{2} \right)\psi_{E a_r b_r} \left( x- \frac{\xi}{2} \right) \mbox{exp} \left( -\frac{i\xi p}{\hbar}  \right) \right).
 \ee
It may be different from $0$ exclusively for
$x \in \left( \frac{a_l+a_r}{2}, \frac{b_l+b_r}{2} \right).$ This interval in general is not contained in the sum of intervals $(a_l, b_l) \cup (a_r, b_r).$
Hence the interference part of the Wigner function (\ref{16.1}) can be nonzero only at points with abscissae, at which the wave functions $\psi_{E a_l b_l}(x)$ and $\psi_{E a_r b_r}(x)$ disappear. The function $W_{E \, int}(x,p)$ is real and it does not contribute to the spatial density of probability, because
 \be
 \label{19}
\varrho_{int}(x)= \int_{-\infty}^{+\infty} dp \,W_{E \, int}(x,p)=0.
\ee
 Therefore
 \[
 \int_{-\infty}^{+\infty}
 dx \int_{-\infty}^{+\infty} dp \,W_{E \, int}(x,p)=
 \int_{\frac{a_l+a_r}{2}}^{\frac{b_l+b_r}{2}} dx \int_{-\infty}^{+\infty} dp \,W_{E \, int}(x,p)=0.
 \]
 The following integrals also vanish
 \[
 \int_{-\infty}^{+\infty} dx \int_{-\infty}^{+\infty} dp \,W_{E \, int}(x,p) W_{E a_l b_l}(x,p) =
 \int_{-\infty}^{+\infty} dx \int_{-\infty}^{+\infty} dp \,W_{E \, int}(x,p) W_{E a_r b_r}(x,p) =0.
 \]
  Both limits of the integral (\ref{18}) with respect to $d \xi$ are always negative and the length of the interval of integration fulfills the inequality
  \[
 \Big| {\rm Max.}[2(a_l-x),2(x-b_r)] - {\rm Min.}[2(b_l-x),2(x-a_r)] \Big| \leq {\rm Min.}[2(b_l-a_l),2(b_r-a_r)].
  \]

For any observable $A(x)$ depending only on the position $x$,
the interference Wigner function $\,W_{E \, int}(x,p)$ does not influence the mean value of  $A(x),$ because
\[
 \int_{-\infty}^{+\infty} dx \int_{-\infty}^{+\infty} dp \,W_{E \, int}(x,p) A(x) =0.
 \]

The existence of the interference Wigner function leads to an apparent paradox. Indeed, assume that a wave function, which is not necessarily an energy eigenfunction, is of the form
\be
\label{20}
\psi(x)= Y(x-a_1)\psi_{a_1b_1}(x)Y(b_1-x) + Y(x-a_2)\psi_{a_2 b_2}(x)Y(b_2-x),
\ee
\[
-\infty < a_1 < b_1 < a_2 < b_2 < \infty.
\]
Thus this function vanishes between points $b_1$ and $a_2.$
Therefore the density of probability of detection of the system in any spatial point from the interval $[b_1,a_2]$ is zero. From our previous considerations we deduce, that  for  any observable $A(x), \; \left(\frac{\partial A(x)}{\partial p} \right)_x=0$ with the support being a strip  in the spatial interval $[b_1,a_2],$ the mean value $\Big< A(x)\Big>=0$ although the Wigner function $W(x,p)$ of this state is different from zero there.

\begin{figure}[h]\centering
\includegraphics[scale=.9]{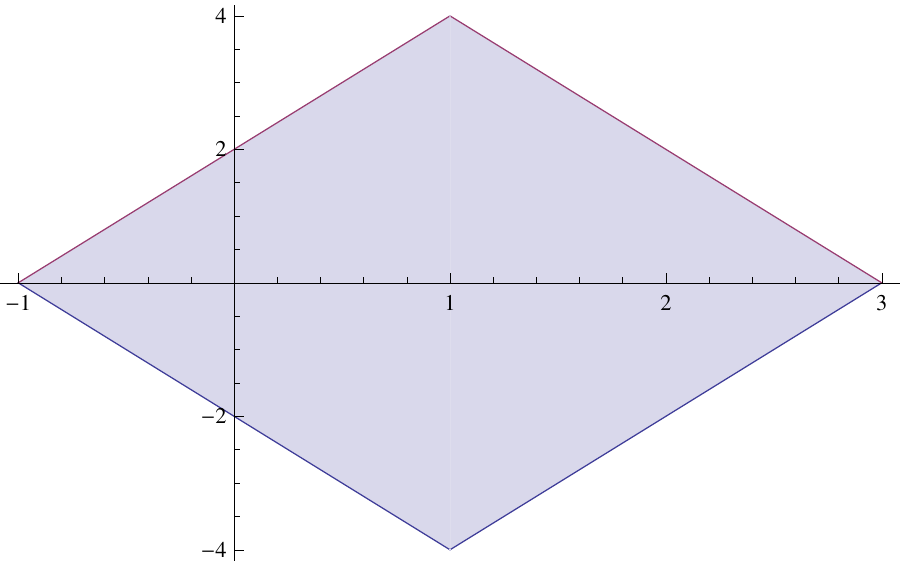}
\caption{The maximal support of the function $f(x,\xi)$ in the coordinates $(x, \xi)$ for $b_1=-1, \; a_2=3.$}
\label{pic0.1}
\end{figure}

Even a stronger conclusion can be formulated. Let $A(x,p)$ be a real function defined as
\be
\label{21}
A(x,p)= \left\{\begin{array}{cc}
\int_{{\rm Max.}[2(b_1-x),2(x-a_2)]}^{{\rm Min.}[2(x-b_1),2(a_2-x)]} \,d \xi \,f(x,\xi) \mbox{exp} \left( -\frac{i\xi p}{\hbar}  \right) & {\rm for}\;\; b_1 < x <a_2 \\
 & \\
 0 & {\rm for}\;\; x \leq b_1 \; {\rm and }\; x \geq a_2.
\end{array}
\right.
\ee
The function $f(x,\xi)$ is chosen in such a way that it ensures reality of the function $A(x,p)$ and the existence of the integral in the formula (\ref{21}) but otherwise arbitrarily (see FIG. \ref{pic0.1}).
Then for any state as in (\ref{20}) characterised by  its respective  Wigner function $W(x,p)$ one has that
\[
 \Big< A(x,p) \Big>= \int_{- \infty}^{\infty}dx \int_{- \infty}^{\infty}dp \,W(x,p) A(x,p)=0.
\]

The crucial role of the interference component of a Wigner function will be illustrated in the case of  the ground state of a $1$--D harmonic oscillator.
The wave function of this state is
\be
\label{op1}
\psi_{0}(x)= \left( \frac{M \omega}{ \pi \hbar}\right)^{1/4} \exp \left(- \frac{M \omega x^2}{2 \hbar} \right)\;, \;\;\; E= \frac{\hbar \omega}{2}
\ee
and it can be written as the sum
$
\psi_{0}(x)=\psi_{0(-)}(x)+ \psi_{0(+) }(x)
$
with
\[
\label{op2}
\psi_{0(-) }(x)=
 \left( \frac{M \omega}{ \pi \hbar}\right)^{1/4} \exp \left(- \frac{M \omega x^2}{2 \hbar} \right)Y(-x)\; , \;
 \psi_{0(+) }(x)= \left( \frac{M \omega}{ \pi \hbar}\right)^{1/4} \exp \left(- \frac{M \omega x^2}{2 \hbar} \right)Y(x).
\]
On the other hand  the Wigner eigenfunction of the ground state of the  $1$--D harmonic oscillator is given by
\be
\label{op3}
W_0(x,p)= \frac{1}{\pi \hbar} \exp \left(- \frac{p^2 + M^2 \omega^2 x^2}{ \hbar M \omega} \right),
\ee
which is of the Gaussian type and therefore nonnegative (see \cite{ja2}).

\begin{figure}[H]\centering
\subfloat[The complete Wigner eigenfunction]{\label{pic1}
\includegraphics[width=0.3\textwidth]{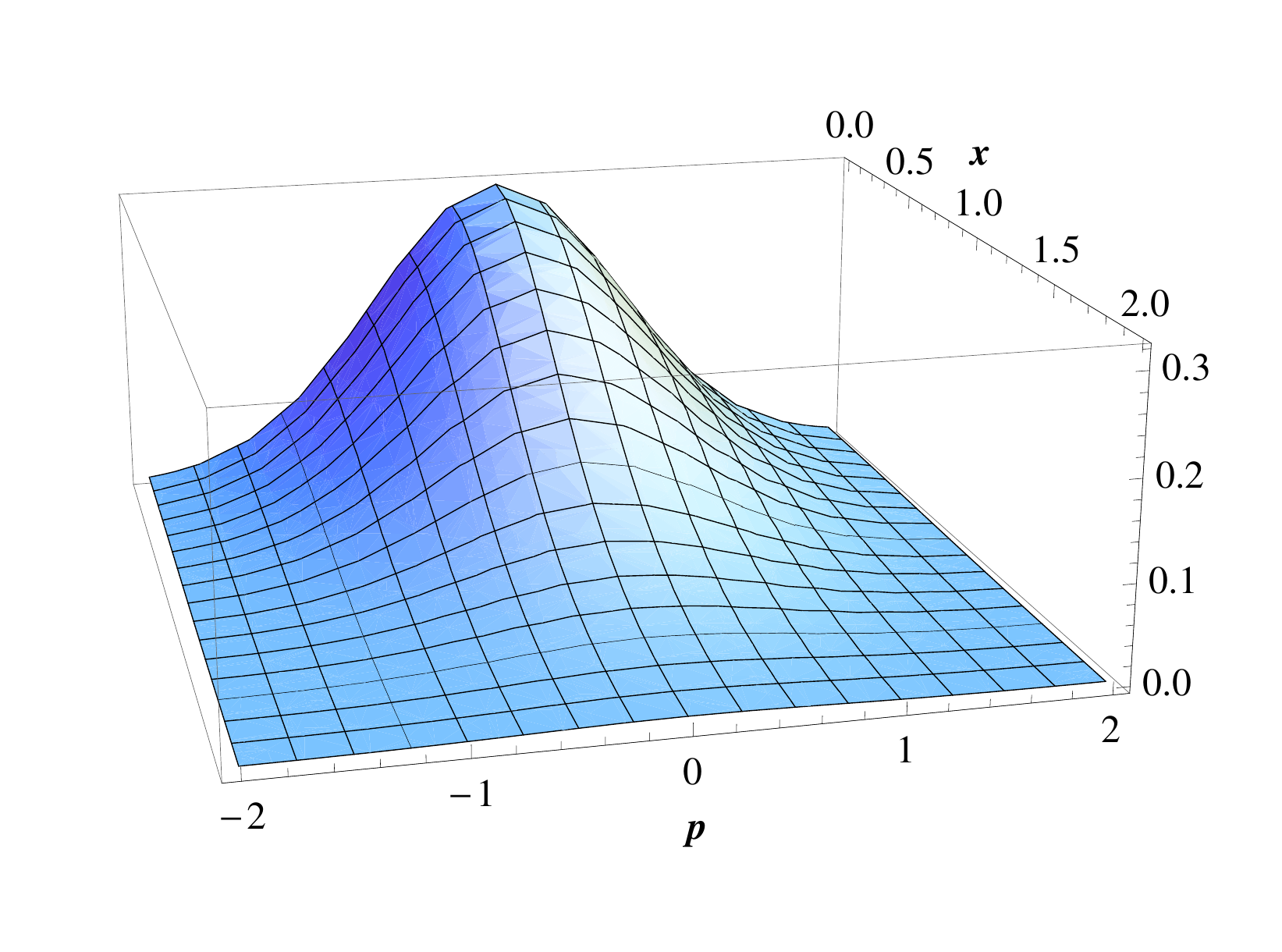}}
\quad
\subfloat[The  Wigner energy eigenfunction without the interference contribution]{\label{pic1.0}
\includegraphics[width=0.3\textwidth]{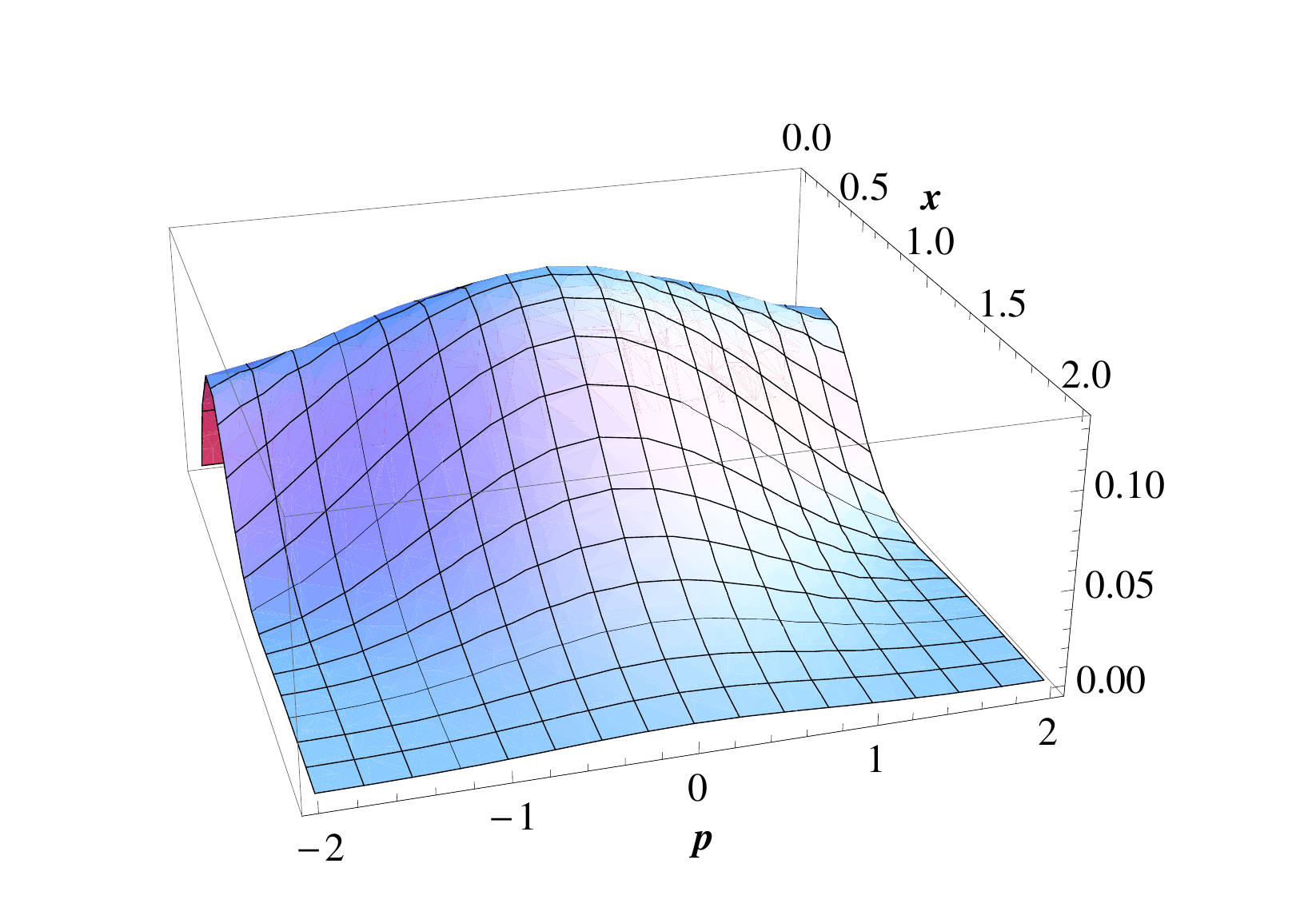}}
\quad
\subfloat[The interference Wigner eigenfunction]{\label{pic1.1}
\includegraphics[width=0.3\textwidth]{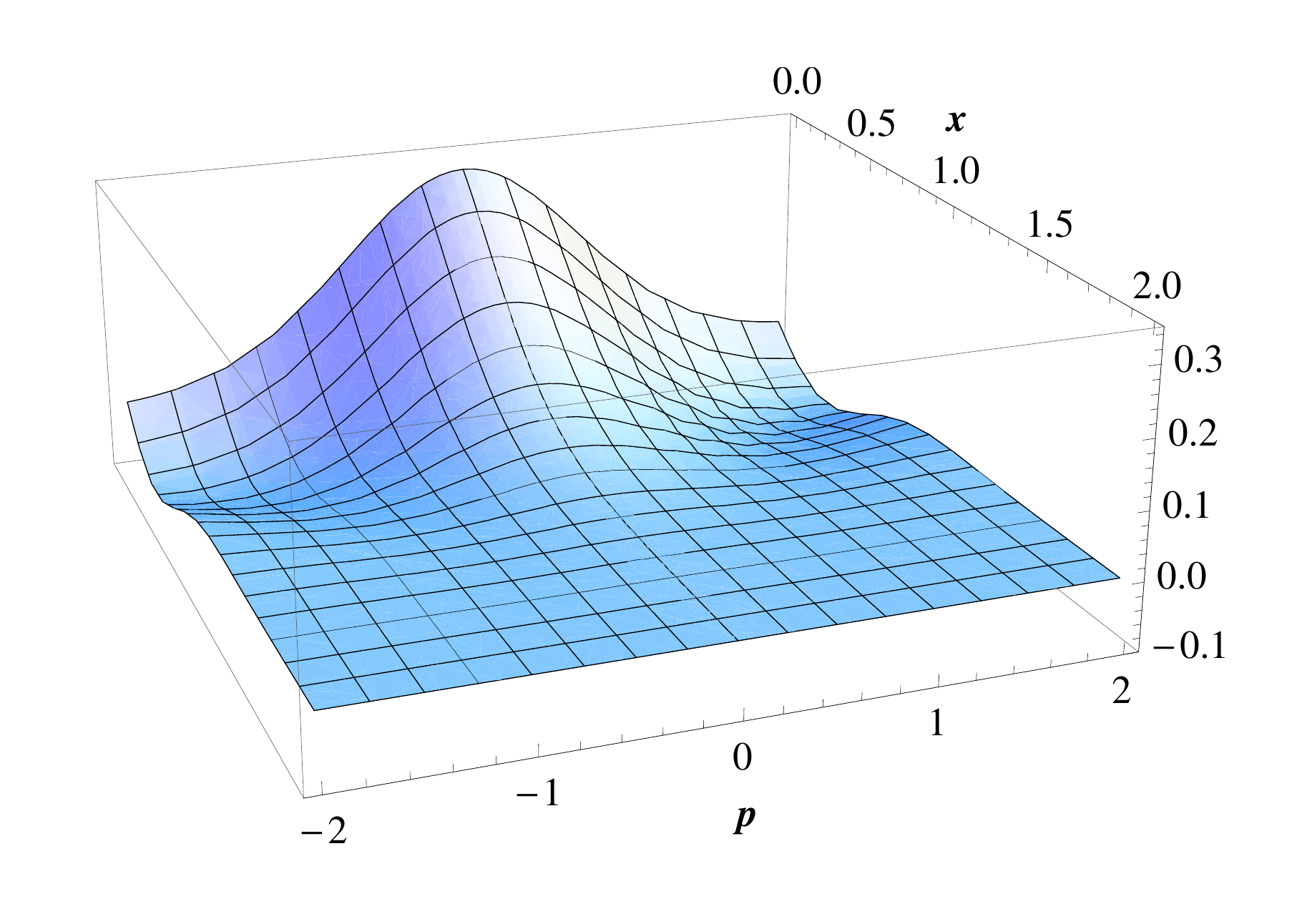}}
\caption{Components of the Wigner eigenfunction in the area $[0, + \infty) \times {\mathbb R}$ for the ground state of the $1$--D harmonic oscillator.}
\label{pic1111}
\end{figure}

The figures illustrating: the complete Wigner function  (\ref{op3}) in the area $[0, + \infty) \times {\mathbb R},$  the Wigner function corresponding to the wave function $\psi_{0(+) }(x)$ and the interference Wigner function between $\psi_{0(-) }(x)$ and $\psi_{0(+)}(x)$ restricted to the region $[0, + \infty) \times {\mathbb R}$ are presented at FIGS. \ref{pic1}, \ref{pic1.0} and \ref{pic1.1}.

For $x=0$ the  component of the Wigner eigenfunction without the interference part disappears. This observation is in agreement with the formula (\ref{16}) applied with the parametres $a_l=0 , \; b_l= \infty.$

Although the complete ground state Wigner  eigenfunction is positive,  its interference component admits negative values as can be appreciated at FIG. \ref{pic1.1}.

The Wigner function related to the wave function (\ref{15.3}) represented by a sum of spatially separable functions can be also found in another way. For every function $\psi_{E a_l b_l}(x)= Y(x-a_l) \psi_{E a_l b_l}(x) Y(b_l-x)$ we introduce a new function
\[
\tilde{\psi}_{E a_l b_l}(x):= Y(a_l-x) +  Y(x-a_l) \psi_{E a_l b_l}(x) Y(b_l-x) + Y(x-b_l),
\]
which is not square integrable, but it is a tempered distribution.  Now the complete energy eigenfunction is the product of functions
\[
\psi_E(x)= \prod_{l=1}^k \tilde{\psi}_{E a_l b_l}(x)\; ,\; - \infty \leq a_1 < b_1=a_2 < b_2=a_3 < \ldots <b_{k-1}= a_k < b_k \leq \infty.
\]
Therefore the Wigner energy eigenfunction can be calculated as the convolution (\ref{15}) of Wigner functions representing  $\tilde{\psi}_{E a_l b_l}(x).$ However, these Wigner functions are usually tempered generalised functions and are considered at Appendix \ref{appA}.

\subsection{Compatibility conditions for Wigner functions}

In the previous subsection the problem of relations between partial Wigner functions defined at spatially separable intervals has been discussed. Now we propose a way in which these partial functions can be put together. The lacking elements are coefficients standing at them. There are a few ways, in which they can be found. The most natural seems to this one based on considerations devoted to an analogous problem in Sec. \ref{sec3}.

First in each region we solve the equation for phase. For the intervals, at which  the series expansion (\ref{6}) holds, this is the system (\ref{7}). Close to turning points it is necessary to deal with the Schroedinger equation for a polynomial potential. Then from the requirement of continuity for the wave function we obtain factors standing at partial wave functions. These factors are transferred to the partial Wigner functions and interference Wigner functions with the use of the Weyl correspondence (\ref{12}).

\section{Examples of the  WKB construction in deformation quantization}

\label{ostatni}

\setcounter{equation}{0}

This section contains two examples of application of the WKB method proposed before to $1$--D energy eigenvalue problems in deformation quantization. The first one corresponds to the bound states of a $1$--D harmonic oscillator and the second to the unbound states of a Poeschl -- Teller potential. Since these two problems can be solved strictly they provide us the possibility to compare exact results with the ones obtained in the first order of the WKB approximation.

\subsection{The WKB approximated Wigner energy eigenstates for the $1$--D harmonic oscillator}

The Hamilton function of the $1$--D harmonic oscillator is given by the formula
\be
\label{46}
H(x,p)= \frac{p^2}{2M} + \frac{M \omega^2 x^2}{2}.
\ee
 Applying the energy  quantization rule (\ref{45}) we obtain that energy levels are determined  by the relation
 $
  E_n= \hbar \omega \left(n+\frac{1}{2} \right), n=0,1,2,\ldots
  $
  exactly  like in the strict procedure. The corresponding turning points are given by
$
-x_0= - \sqrt{\frac{(2n+1)\hbar}{M \omega}} $  and $ x_0=\sqrt{\frac{(2n+1)\hbar}{M \omega}}.
$

Then combining the applicability condition (\ref{8}) and a demand of compatibility of a linear approximation of the potential $ \frac{M \omega^2 x^2}{2}$ with this real potential   near to the turning points we see, that the quasi--classical method can be definitely applied for energy eigenvalues fulfilling the condition $n \geq 8.$

Moreover, we divide the domain ${\mathbb R}$ of Eq. (\ref{5}) for the harmonic oscillator into five regions:
\begin{enumerate}
\item
the classically forbidden areas $
A_L= \left(- \infty,-\frac{5}{4}x_0 \right]$ and $A_R=\left[ \frac{5}{4}x_0, \infty \right)$ in which the series expansion (\ref{6}) works,
\item
the intervals  $B_L= \left[-\frac{3}{2}x_0,-\frac{3}{4}x_0 \right]$ and  $B_R=\left[\frac{3}{4}x_0,\frac{3}{2}x_0\right]$ containing the turning points
\item
and
the classically accessible part  $ C= \left[-\frac{7}{8}x_0, \frac{7}{8}x_0\right].$
\end{enumerate}
The notation used above is in agreement with the one proposed for FIG. \ref{pic1.00}.

In the areas $A_L, C$ and $A_R$ we solve Eqs. (\ref{7}) up to the first order approximation. In the regions $B_L$ and $B_R$  we deal with the Schroedinger equation with a linear potential.

A list of phases and normalising factors for the five enumerated regions is presented below, where for simplicity we have  omitted the index `$(1)$' at  normalising factors.
\begin{enumerate}
\item
In the interval $A_L$ it is found that
\[
\sigma_0=  \frac{i M \omega}{2} \left\{ -x\sqrt{x^2- \frac{\hbar (2n+1)}{M \omega}} -
\frac{\hbar (2n+1)}{M \omega} \ln \left(\frac{-x+ \sqrt{x^2- \frac{\hbar (2n+1)}{M \omega}}}{\sqrt{\frac{\hbar (2n+1)}{M \omega}}} \right)
\right\},
\]
\[
\sigma_1= - \frac{1}{4} \ln \left( M^2 \omega^2 x^2 - M \hbar \omega (2n+1) \right),
\]
and
$
D_{A_L}= \frac{N}{2},
$
where $N$ is a parameter determined by the normalisation requirement.
\item
In the region $B_L$ the series representation of the phase $\sigma$ does not work. Since  the potential has  been approximated there by a linear function, we see that
\[
\sigma= -i \hbar \ln \left( \Phi \left[- 2^{1/3}\sqrt{\frac{M \omega}{\hbar}} (2n+1)^{1/6} \left(\sqrt{\frac{\hbar (2n+1)}{M \omega}}+x \right) \right]  \right)
\]
and the normalising coefficient $C_{B_L}= \frac{N}{2^{1/6 }M^{1/4}\omega^{1/4} \hbar^{1/4}(2n+1)^{1/12}}.$
\item
In the classically accessible area $C$ there are two physically acceptable solutions. As it was shown in Sec. \ref{sec3}, they differ at signs at $\sigma_0$ but they have the same phases $\sigma_1.$ Thus
\[
\sigma_{0\,I}=-\sigma_{0\,II}=
 \]
 \[
    \frac{(2n+1)\hbar}{2}  \left[ x \sqrt{\frac{M \omega}{(2n+1 )\hbar}} \sqrt{1-\frac{M\omega x^2}{(2n+1) \hbar}} +\arcsin\left(x \sqrt{\frac{M\omega }{(2n+1) \hbar}} \right) \right] + \frac{(2n+1) \pi}{4}
\]
and
\[
\sigma_{1\,I}=\sigma_{1\,II}= - \frac{1}{4} \ln \left( M \hbar \omega (2n+1)- M^2 \omega^2 x^2  \right).
\]
The normalising factors are $D_{I\,C}=\frac{N(1-i)\sqrt{2}}{4}$ and  $D_{II\,C}=\frac{N(1+i)\sqrt{2}}{4}.$
\item
In the region $B_R$ one gets
\[
\sigma= -i \hbar \ln \left( \Phi \left[ - 2^{1/3}\sqrt{\frac{M \omega}{\hbar}} (2n+1)^{1/6} \left(\sqrt{\frac{\hbar (2n+1)}{M \omega}}-x \right) \right] \right)
\]
with the factor $C_{B_R}= \frac{(-1)^n N}{2^{1/6 }M^{1/4}\omega^{1/4} \hbar^{1/4}(2n+1)^{1/12}}.$
\item
Finally in the interval $A_R$
\[
\sigma_0=  \frac{i M \omega}{2}  \left\{ x\sqrt{x^2- \frac{\hbar (2n+1)}{M \omega}} -
\frac{\hbar (2n+1)}{M \omega} \ln \left(\frac{x+ \sqrt{x^2- \frac{\hbar (2n+1)}{M \omega}}}{\sqrt{\frac{\hbar (2n+1)}{M \omega}}} \right)
\right\},
\]
\[
\sigma_1= - \frac{1}{4} \ln \left( M^2 \omega^2 x^2 - M \hbar \omega (2n+1) \right)
\]
and the coefficient
$
D_{A_R}= \frac{(-1)^n N}{2}.
$

\end{enumerate}
Since the explicit formulas for partial and interference Wigner functions obtained via the Weyl correspondence (\ref{12.10}) are extremely complicated,
 we present the results by the following pictures obtained numerically.

\begin{figure}[H]\centering
\subfloat[The first WKB approximation]{\label{pic1.10}
\includegraphics[width=0.45\textwidth]{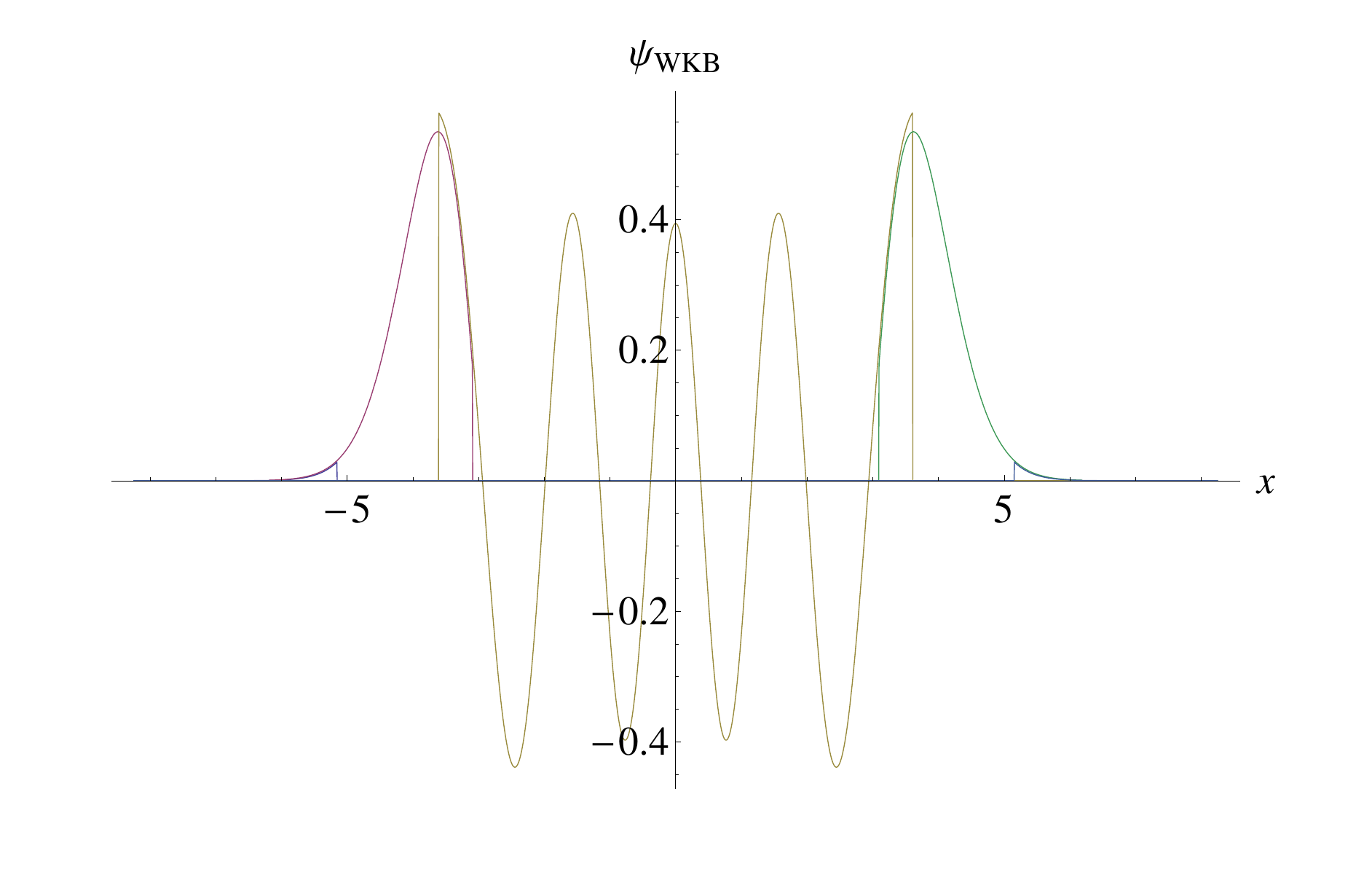}}
\quad
\subfloat[The strict   wave function]{\label{pic1.100}
\includegraphics[width=0.45\textwidth]{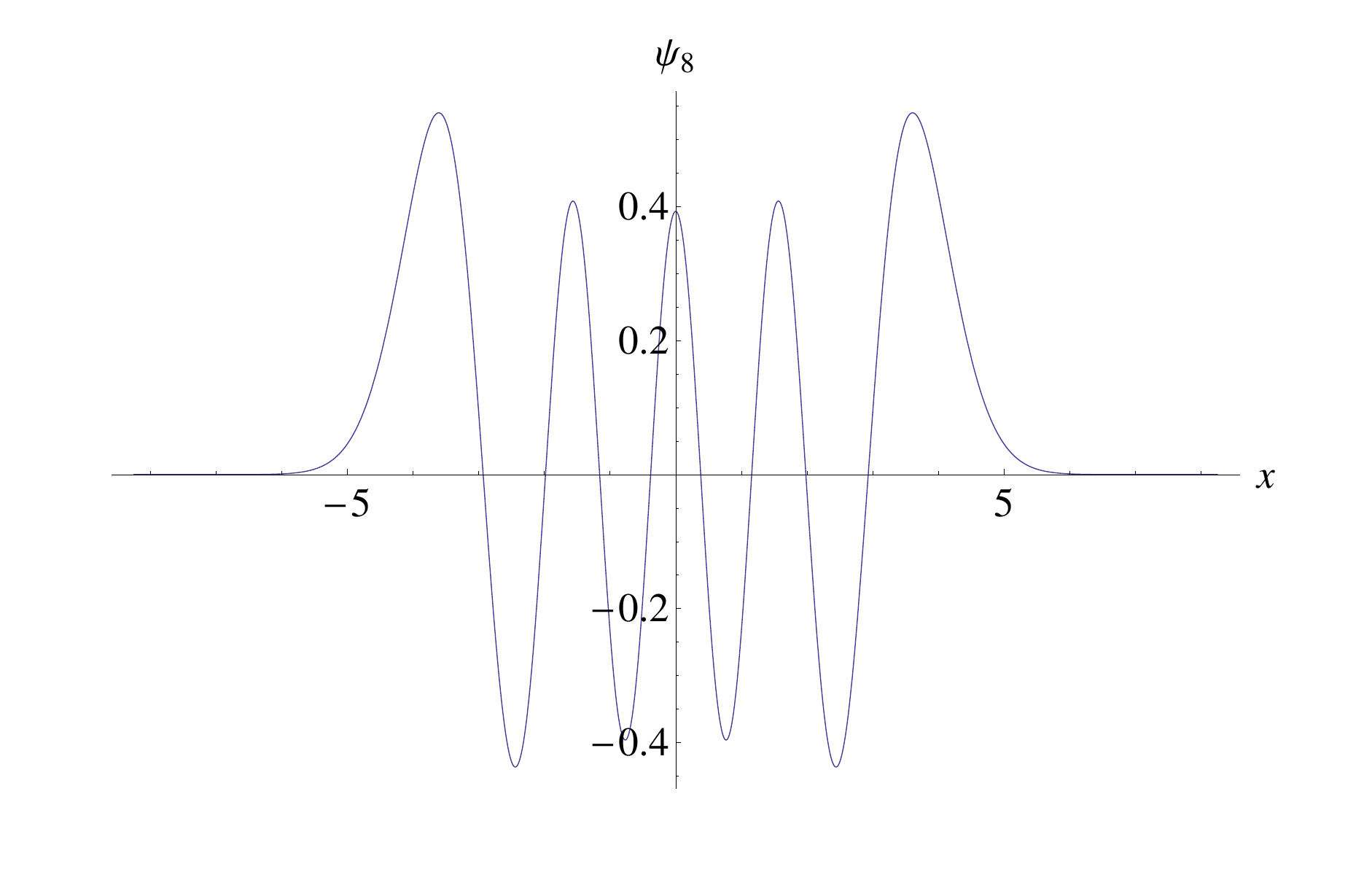}}
\caption{The harmonic oscillator wave energy eigenfunctions in the first WKB approximation and in the strict version for $n=8$. }
\end{figure}

As it can be seen comparing FIG. \ref{pic1.10} and  FIG. \ref{pic1.100} for $n=8,$ the first order of the WKB approximation applied to the  wave energy eigenfunction
generates a function which is in a good agreement with the strict solution.

\begin{figure}[H]\centering
\includegraphics[scale=.65]{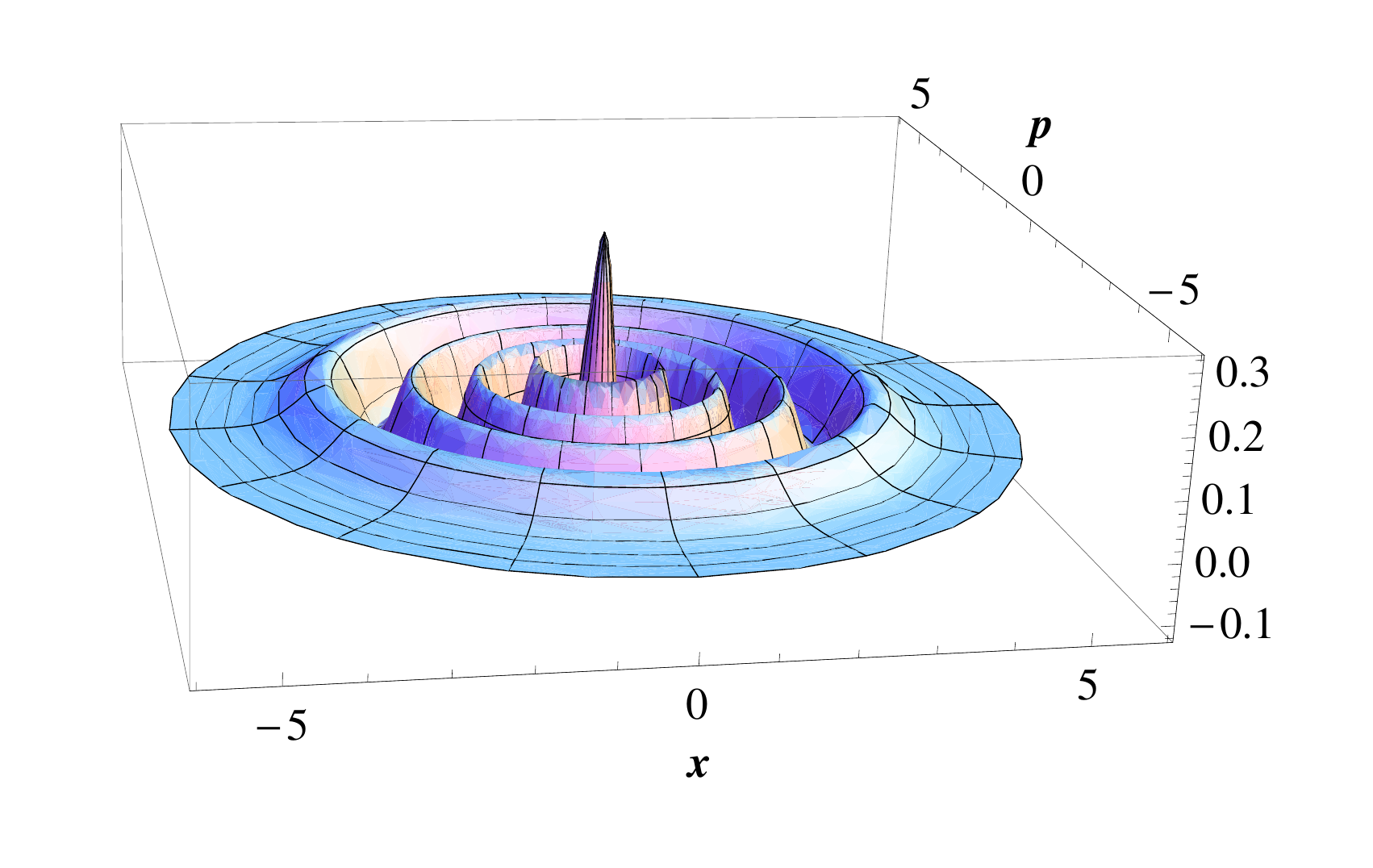}
\caption{The strict Wigner energy eigenfunction for $n=8$.}
\label{pic1.11}
\end{figure}

An analogous situation takes place for the Wigner energy eigenfunction for $n=8.$
 The strict Wigner energy eigenfunction and the Wigner energy eigenfunction in the first order of the WKB approximation look practically the same.
 The exact solution,
  determined by the expression
\be
\label{52}
W_{\hbar \omega (n+1/2)}(x,p)= \frac{(-1)^n}{\pi \hbar} \exp \left(- \frac{2H}{\hbar \omega} \right)L_n \left( \frac{4H}{\hbar \omega}\right),
\ee
where $H$ denotes the Hamilton function (\ref{46}) and $L_n(y):= \sum_{m=0}^n (-1)^m
\left( \begin{array}{c}
n \\n-m
\end{array}
\right) \frac{y^m}{m!}
$  is the $n$th Laguerre polynomial,
  can be seen at FIG. \ref{pic1.11}.

\begin{figure}[H]\centering
\includegraphics[scale=1.0]{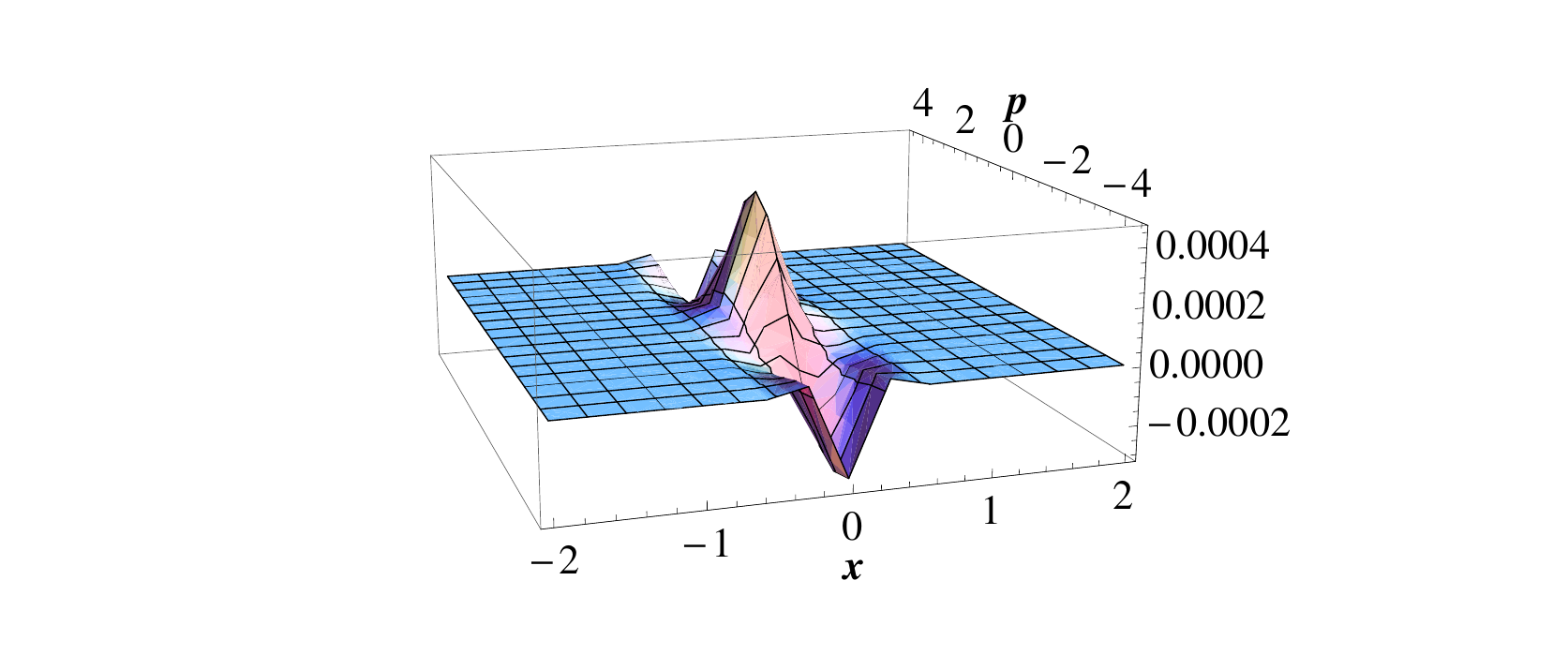}
\caption{The contribution due to the interference between the wave functions from the regions $ A_L$ and $ A_R$ to the Wigner function.}
\label{pic1.13}
\end{figure}

The effect of the interference terms on the complete Wigner energy eigenfunction can be appreciated at FIG. \ref{pic1.13}. Values of this Wigner interference function are in general small comparing to the complete Wigner function but at some points the interference contribution is significant.

\begin{figure}[H]\centering
\includegraphics[scale=.8]{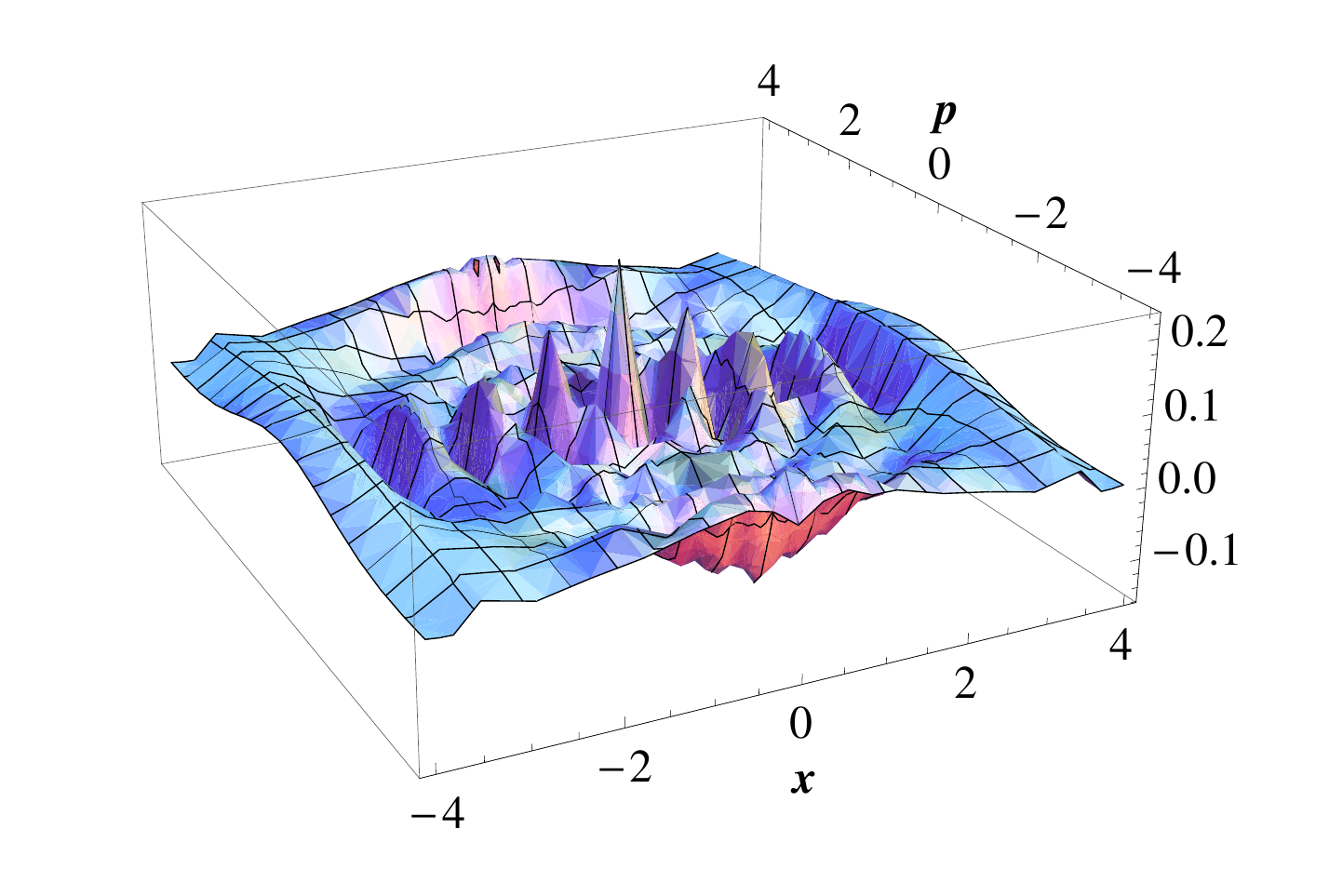}
\caption{The Wigner energy eigenfunction without the interference  contribution.}
\label{pic1.14}
\end{figure}

The Wigner energy eigenfunction without any interference  contribution is presented at FIG. \ref{pic1.14}. This incomplete function is in fact sufficient to deal with the spatial probability.

To complete our analysis it is important to mention the following remarks about  the $0$th approximation of the Wigner energy eigenfunction. In this case the phase $\sigma_0$ is a
solution of the classical Hamilton -- Jacobi equation. However, the class of admissible solutions is larger than in the classical physics, because we accept also
imaginary phases in classically forbidden regions.

\begin{figure}[H]\centering
\includegraphics[scale=0.8]{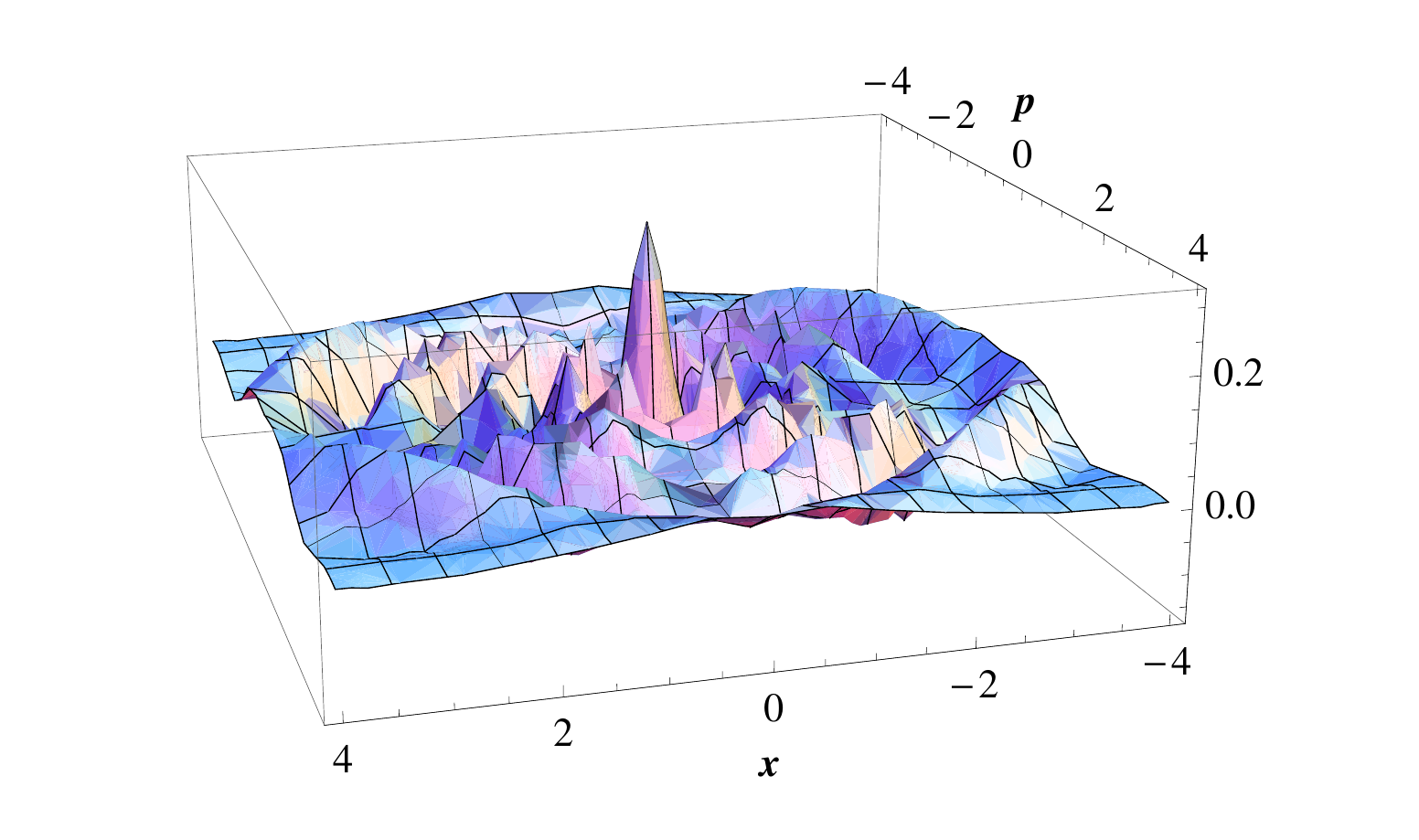}
\caption{The Wigner energy eigenfunction for $n=8$ in the zeroth WKB approximation.}
\label{pic1.15}
\end{figure}

 The Wigner energy eigenfunction in the zeroth WKB approximation for the $n=8$ is presented at FIG. \ref{pic1.15}. In general it is similar to the strict Wigner energy eigenfunction presented at FIG.
 \ref{pic1.11}. Since it takes also negative values, it does not represent a probability distribution and cannot be treated in a classical way.

\subsection{The unbound energy eigenstates of the Poeschl -- Teller potential in deformation quantization}

\begin{figure}[H]\centering
\includegraphics[width=0.5\textwidth]{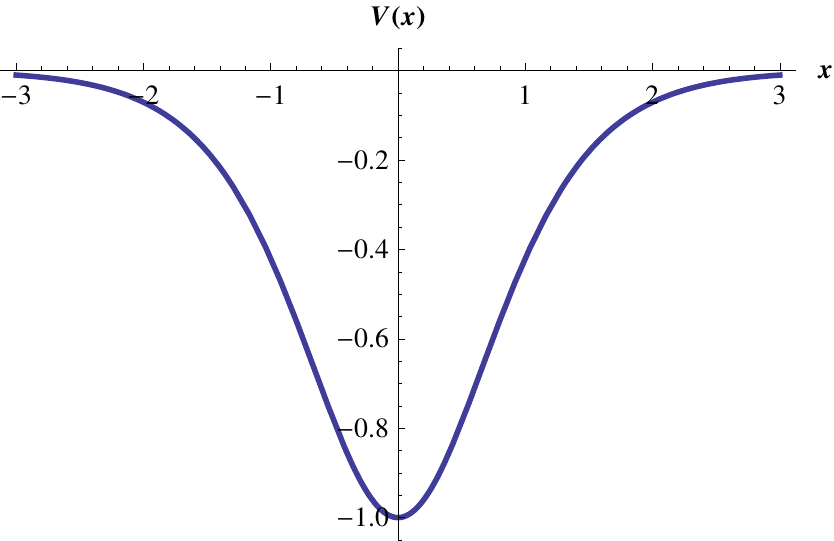}
\caption{The Poeschl -- Teller potential for the parametres $\hbar=a=M=1.$}
\label{pic1.16}
\end{figure}

\begin{figure}[H]\centering
\subfloat[The spatial density of probability]{\label{pic1.17}
\includegraphics[width=0.45\textwidth]{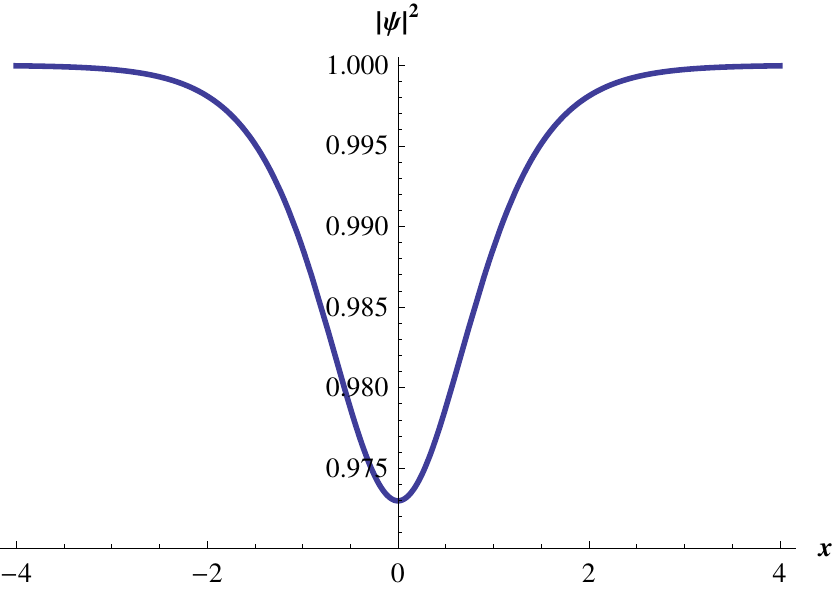}}
\quad
\subfloat[The momenta density of probability]{\label{pic1.180}
\includegraphics[width=0.45\textwidth]{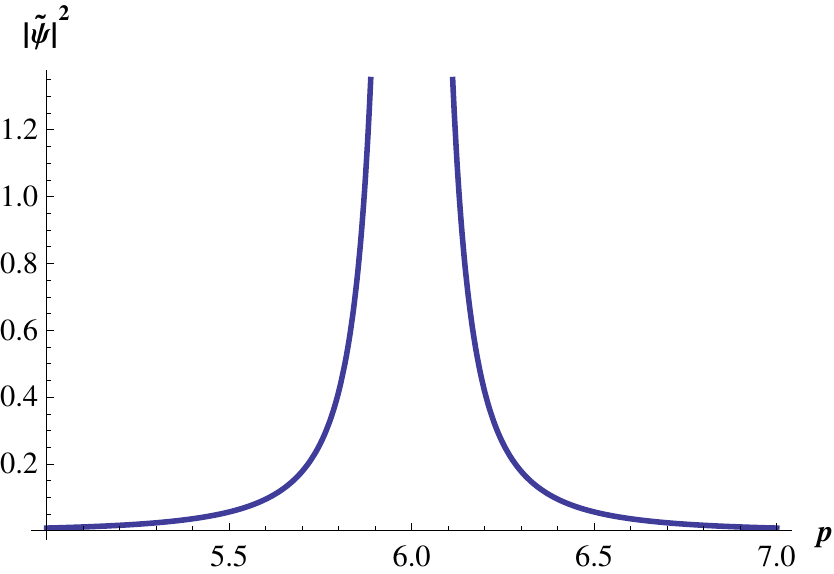}}
\caption{The densities of probability for the unbound state for $k=6$ and $M=a=1$ in the Poeschl -- Teller potential.}
\end{figure}

As the second example we consider a case of unbound states, for which the series expansion (\ref{6}) is valid everywhere.  The chosen potential is
the Poeschl -- Teller potential  described by the expression
\be
\label{m1}
V(x)= - \frac{\hbar^{2}a^{2}}{M} \frac{1}{\cosh^{2}(ax)},
\ee
where $a>0$ is a parameter. The form of this potential is shown at FIG. \ref{pic1.16}.
 The energy eigenvalue problem for this potential is solvable for any positive energy $E>0$ and its solution is the function
\be
\label{m2}
\psi_k(x)= A \left( \frac{ik- a \tanh(ax)}{ik+a}\right)\exp(ikx)\;\;, \;\; k=\sqrt{\frac{2ME}{\hbar}}.
\ee
An interesting feature of the potential (\ref{m1}) is that every incident particle, regardless of its energy, passes right through.
The wave function (\ref{m2}) is not normalisable and the relative spatial density of probability determined by it is illustrated at FIG. \ref{pic1.17}.

In the momentum representation the wave function $\tilde{\psi}_k(p)$ is a sum of two generalised functions. Indeed,
\be
\label{m3}
\tilde{\psi}_k(p)= A \sqrt{2 \pi \hbar} \frac{ik}{ik+a} \left(\delta(p-k \hbar)-
\frac{ 1}{2 k \hbar}\,{\rm vp}\frac{1}{\sinh\left( \frac{\pi(p-k \hbar)}{2 a \hbar}\right)} \right),
\ee
where ${\rm vp}$ denotes the principal value of the function.
 The density probability with respect to momenta is not defined but the relative probability is illustrated at FIG. \ref{pic1.180}.

In the quasi -- classical approximation the series expansion (\ref{6}) can be applied everywhere. Moreover, the Poeschl -- Teller potential is reflectionless and it is assumed that a source of particles is localised at $- \infty$. As the limit in the integral determining the phase $\sigma_0$ we put $x_0=0.$ Thus
\[
\sigma_0= \frac{\hbar \sqrt{k^2 \cosh^2 ax + 2a^2}}{\sqrt{k^2 \cosh 2ax + 4a^2+k^2}}
\left[2 \arctan \left(\frac{2a \sinh ax}{\sqrt{k^2 \cosh 2ax + 4a^2+k^2}} \right) +
\right.
 \]
 \setcounter{orange}{1}
\renewcommand{\theequation} {\arabic{section}.\arabic{equation}\theorange}
 \be
 \label{m100}
 \left.
    \frac{k}{a} {\rm arcsinh}\, \left(\frac{k \sinh ax}{\sqrt{2a^2+k^2}} \right) \right]
 \;\;\; {\rm and}
\ee
\addtocounter{orange}{1}
\addtocounter{equation}{-1}
\be
\label{m101}
\sigma_1=-\frac{1}{2} \ln \left( \hbar \cosh ax \sqrt{k^2 \cosh^2 ax + 2a^2} \right).
\ee
\renewcommand{\theequation} {\arabic{section}.\arabic{equation}}
Since the integral (\ref{12.10}) is divergent in both:  the strict (\ref{m2})  and the approximated (\ref{m100}, \ref{m101}) case, to calculate respective Wigner functions is necessary to apply the formula (\ref{a8}). Thus to show that the WKB method leads to a reasonable result, we compare  the wave function (\ref{m2}) and the approximated  function $2.5 \cdot \exp \left(\frac{i}{\hbar} \sigma_0 + \sigma_1 \right).$ The values of the approximated  function have been rescaled by the factor $2.5$ because  the most natural choice of the free term equal to $0$ in $\sigma_0$ gives a value of the factor  $|A| \neq 1$ in the formula (\ref{m2}).
\begin{figure}[H]\centering
\subfloat[The real parts]{\label{pic1.19}
\includegraphics[width=0.45\textwidth]{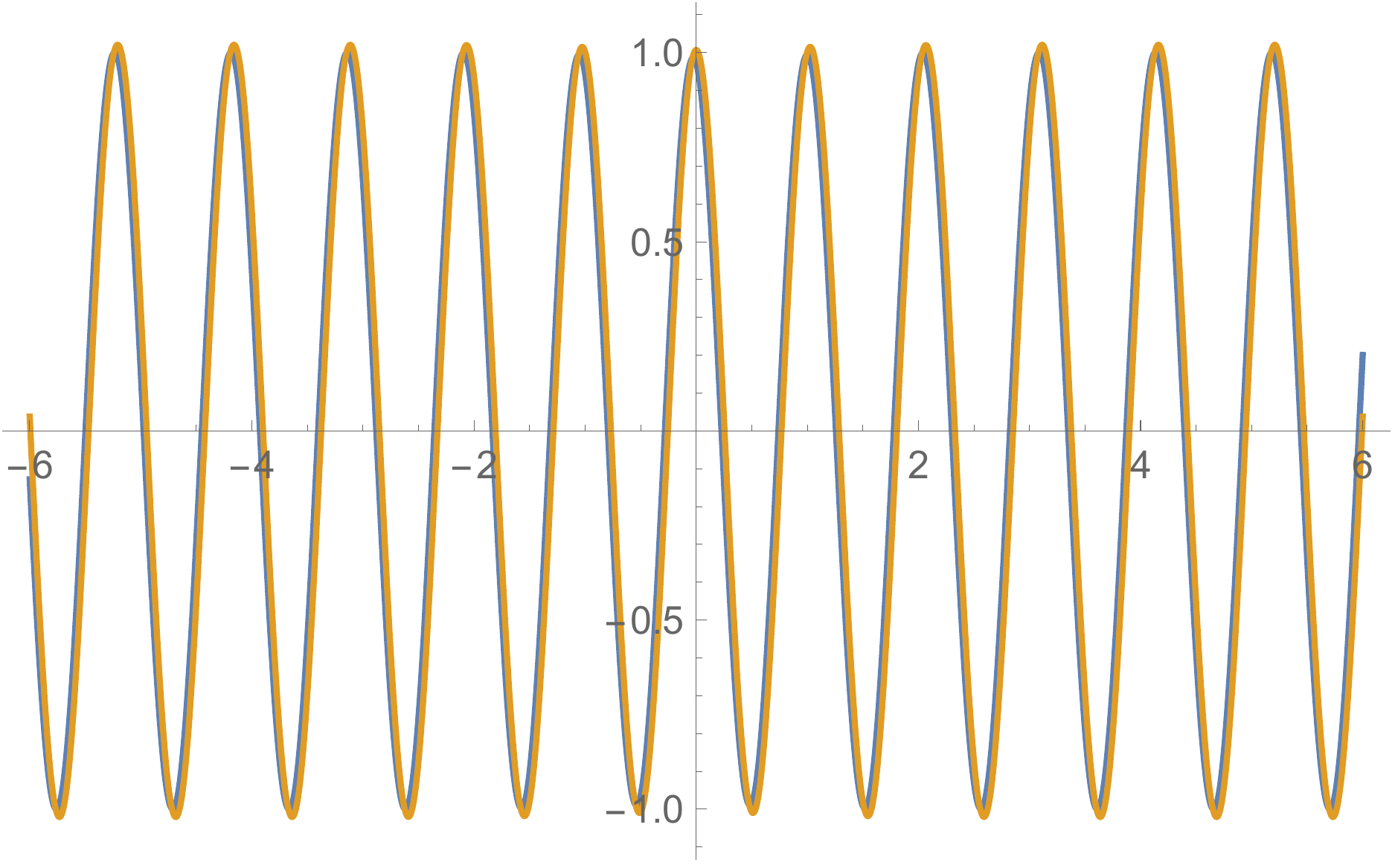}}
\quad
\subfloat[The imaginary parts]{\label{pic1.190}
\includegraphics[width=0.45\textwidth]{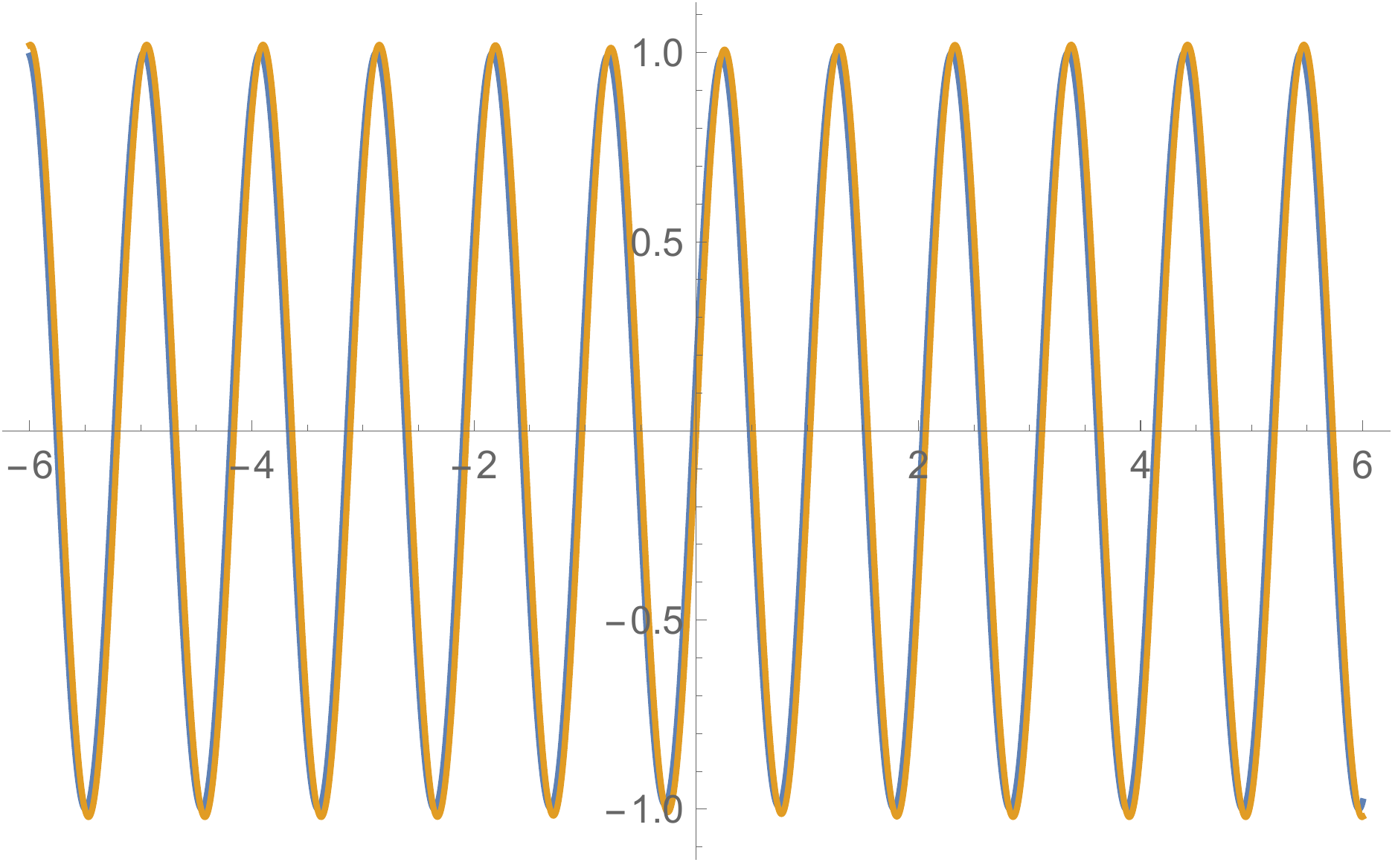}}
\caption{Comparison between the real and imaginary parts of the strict wave function and its first WKB approximation for the unbound state $k=6,\; M=a=1.$}
\label{pic1.1900}
\end{figure}

The results can be appreciated at FIGS. \ref{pic1.19} and \ref{pic1.190}, where
the strict function is in blue and the approximated one in yellow. It is hard to distinguish between them.
The respective  Wigner function can be found at the end of Appendix \ref{appB} at FIG. \ref{pic1.18}.

\section{Conclusions}
\setcounter{equation}{0}

In this work a careful analysis of obtaining an approximate Wigner function by means of WKB approximation of the wave function is presented. In particular, expressions for the Wigner function coming from a product and a superposition of  functions are found. The WKB algorithm adapted to the deformation quantization formalism  enables us to find approximate energy levels and  approximated Wigner energy eigenfunctions.  Moreover, the $n$th approximation of the Wigner energy eigenfunction is determined by the $n$th approximation of the phase $\sigma(x).$

The WKB approximation in deformation quantization is an approach alternative to the wave quantum mechanics version of the quasi -- classical method. It starts from Eq. (\ref{3}) and through formula (\ref{12.10}) leads directly to the Wigner function so there is no step, at which any use of a wave function would be required.

The initial step of the quasi -- classical method is solving of the classical Hamilton -- Jacobi stationary equation.  However it does not mean that in the $0$th approximation we deal with a `classical' limit of the Wigner function. The reason is that in quantum considerations also imaginary solutions of the   Hamilton -- Jacobi  equation are acceptable.

There are some important difficulties when the quasi -- classical method is applied. The principal one arises from the non -- locality of the Weyl correspondence (\ref{12}). Indeed, to find the Wigner energy eigenfunction at an arbitrary point $(x,p)$ of the phase space we have to know the complete phase (phases) $\sigma(x)$ or, equivalently, the complete wave function. Now, since the wave function is in general a sum of spatially separated functions, interference terms are present and for a problem with several spatial regions it is time -- consuming to calculate the complete Wigner function.

Another disadvantage of the proposed algorithm is the fact, that on the contrary to the idea of the Hamilton formalism, the considerations have been done from a spatial point of view. Thus positions and momenta are not treated on an equal footing.
\vspace{1cm}

\noindent{\Large \textbf{Acknowledgments}}

We are grateful to Prof. Maciej Przanowski for his help and valuable remarks.

This work was partially supported by the CONACYT research grant 103478. In addition
R. C. and F. J. T. were partially supported by SNI-M\'exico, COFAA, EDI and the SIP-IPN
grants  20141498, 20144150, 20150975 and 20151031.

\appendix

\section{The Wigner function of a nonnormalisable state}
\label{appA}
\renewcommand{\theequation} {\Alph{section}.\arabic{equation}}

\setcounter{equation}{0}

In the case when a wave function is not square integrable, the integral in formula (\ref{12}) of a Wigner function may not be convergent. Thus it is necessary to propose some extension  of expression  (\ref{12}) applicable to states represented by generalised functions.

To establish notation we quote three definitions:
\begin{de}
Let $\varphi(z)$ be a function from the Schwartz space ${\cal S}.$ Then its {\bf inverse Fourier transform} is determined by the integral
\be
\label{a1}
\tilde{\cal F}[\varphi(z)](t)=
\tilde{\cal F}_z[\varphi](t)
:= \frac{1}{\sqrt{2 \pi}}\int_{-\infty}^{\infty}\varphi(z) \exp(-izt)dz.
\ee
\end{de}
Generalisation of Def. \ref{a1} on functions of many variables is straightforward.

The inverse Fourier transform of a function from the Schwartz space is also an element of ${\cal S}.$ A tempered distribution $T$ is a linear continuous functional over the vector space ${\cal S}.$ The set of tempered generalised functions will be denoted by ${\cal S}'.$
\begin{de}
An inverse  Fourier transform $\tilde{\cal F}[T] $ of a tempered distribution $T$ is a tempered generalised function satisfying the equality
\be
\label{a2}
\Big< \tilde{\cal F}[T](t), \varphi(t)\Big>:= \Big<T,  \tilde{\cal F}[\varphi(z)](t)\Big>
\ee
for every $\varphi \in  {\cal S}. $
\end{de}
\begin{de}
The {\bf convolution} of two tempered generalised functions $S$ and $T$ (if it exists) is a generalised function $S*T$ defined by the condition
\be
\label{a2.1}
\forall\; \varphi \in  {\cal S} \;\;\;\Big<S * T,\varphi \Big>= \Big<S_x \otimes T_y,\varphi(x+y)\Big>.
\ee
\end{de}
The inverse Fourier transform of the product of tempered distributions $S\cdot T$ (if it is well defined) is the convolution of their respective inverse Fourier transforms
\be
\label{a2.2}
\tilde{\cal F}[S \cdot T]= \frac{1}{\sqrt{2 \pi}} \, \tilde{\cal F}[S]* \tilde{\cal F}[T].
\ee

\setcounter{orange}{1}
\renewcommand{\theequation} {\Alph{section}.\arabic{equation}\theorange}
It is well known that the inverse Fourier transform satisfies the following properties:
\begin{enumerate}
\item
a translation of the argument leads to the relation
\be
\label{aa1}
\forall\; a \in {\mathbb R} \;\;\;
\tilde{\cal F}[T(z+a)](t)= \exp(iat) \tilde{\cal F}[T(z)](t),
\ee
\addtocounter{orange}{1}
\addtocounter{equation}{-1}
\item
a multiplication of the argument  by a real number implies the equality
 \be
\label{aa2}
\forall\; a \in {\mathbb R}\setminus \{0\}  \;\;\;
\tilde{\cal F}[T(az)](t)=\frac{1}{|a|} \tilde{\cal F}[T(z)] \left(\frac{t}{a} \right),
\ee
\addtocounter{orange}{1}
\addtocounter{equation}{-1}
\item
the complex conjugation of a tempered generalised function $T$ is transformed into
 \be
\label{aa3}
\tilde{\cal F}[\overline{T}(z)](t)=\overline{\tilde{\cal F}[T(z)](-t)},
\ee
\addtocounter{orange}{1}
\addtocounter{equation}{-1}
\item
the inverse Fourier transform of the derivative
\be
\label{aa4}
\tilde{\cal F} \left[\frac{dT(z)}{dz} \right] (t)= it \tilde{\cal F}[T(z)](t).
\ee
\end{enumerate}
\renewcommand{\theequation} {\Alph{section}.\arabic{equation}}

One  sees that Def. (\ref{12}) of a Wigner function related to a state $\big|\psi \big>$ can be written as
\be
\label{a3}
W(x,p)= \frac{1}{\sqrt{2 \pi} \hbar} \tilde{\cal F}_{\xi} \left[\overline{\psi} \left( x + \frac{\xi}{2} \right)
\psi \left( x - \frac{\xi}{2} \right) \right]   \left(\frac{p}{\hbar}\right),
\ee
where $\xi$ is a variable and $x$ plays a role of a parametre. We do not assume that $\big|\psi \big>$ is an energy eigenstate.

Applying the fact that the inverse Fourier transform of the product of functions is the convolution of inverse Fourier transforms one obtains that
\be
\label{a4}
W(x,p)= \frac{1}{2 \pi \hbar}
\left(
  \tilde{\cal F}_{\xi} \left[\overline{\psi} \left( x + \frac{\xi}{2} \right)\right] *
\tilde{\cal F}_{\xi} \left[ \psi \left( x - \frac{\xi}{2} \right) \right]
\right)
    \left(\frac{p}{\hbar}\right).
\ee
Using properties (\ref{aa1}), (\ref{aa2}) and (\ref{aa3}) one gets
\setcounter{orange}{1}
\renewcommand{\theequation} {\Alph{section}.\arabic{equation}\theorange}
\be
\label{a5}
\tilde{\cal F}_{\xi} \left[\overline{\psi} \left( x + \frac{\xi}{2} \right)\right] \left(t\right)=2 \exp\left(2 itx\right)
\overline{\tilde{\cal F}[\psi(\xi)]\left(-2t\right)}
\ee
\addtocounter{orange}{1}
\addtocounter{equation}{-1}
and
\be
\label{a6}
\tilde{\cal F}_{\xi} \left[ \psi \left( x - \frac{\xi}{2} \right) \right]
    \left(t \right)=2 \exp\left(-2 itx \right) \tilde{\cal F}[\psi(\xi)]\left(-2t \right).
\ee
\renewcommand{\theequation} {\Alph{section}.\arabic{equation}}
Therefore the Wigner function representing a state $\big|\psi \big>$ equals
\be
\label{a7}
W(x,p)= \frac{1}{ \pi \hbar}
\left(\overline{ \exp(it x)\tilde{\cal F}[\psi(\xi)]}(t)*_t \exp(it x) \tilde{\cal F}[\psi(\xi)](t) \right) \left(-\frac{2 p}{\hbar}\right)
\ee
or
\be
\label{a8}
W(x,p)= \frac{1}{ \pi \hbar}
\left(\overline{ \tilde{\cal F}[\psi(\xi+x)]}(t)*_t  \tilde{\cal F}[\psi(\xi+x)](t) \right) \left(-\frac{2 p}{\hbar}\right).
\ee

\section{The Wigner function of an unbound state in the Poeschl --Teller potential}

\renewcommand{\theequation} {\Alph{section}.\arabic{equation}}

\label{appB}

\setcounter{equation}{0}

From the relationship (\ref{a7}) one can see that the starting point for calculating a Wigner function for a state $\big| \psi \big>$ is the inverse Fourier transform of the wave function $\psi(x)$ of this state. In our problem the wave function is expressed as  (\ref{m2}). Direct calculations lead to the conclusion that
\be
\label{B1}
\tilde{\cal F}_x \left[ A \left( \frac{ik- a \tanh(ax)}{ik+a}\right)\exp(ikx) \right](t)= \frac{A  i}{a+ik}\sqrt{\frac{\pi}{2}}
\left(2k \delta(t-k)- {\rm vp}\frac{1}{\sinh\left(\frac{\pi(k-t)}{2a}\right)}
\right).
\ee
Therefore the Wigner  eigenfunction for an eigenvalue $E=\frac{\hbar^2 k^2}{2m}$ equals to the following convolution
\[
W_{\frac{\hbar^2 k^2}{2m}}(x,p)=\frac{|A|^2}{2 \hbar(a^2+k^2)}\left[\left( 2k\exp(-ikx)\delta(t-k)-
{\rm vp}\frac{\exp(-itx)}{\sinh\left(\frac{\pi(k-t)}{2a}\right)}
\right) *_t \right.
\]
\be
\label{B2}
\left. \left(2k\exp(ikx)\delta(t-k)-
{\rm vp}\frac{\exp(itx)}{\sinh\left(\frac{\pi(k-t)}{2a}\right)}
\right) \right]
\left(-\frac{2 p}{\hbar}\right).
\ee
The most complicated part is to derive the convolution
\[
\left[{\rm vp}\frac{\exp(-itx)}{\sinh\left(\frac{\pi(k-t)}{2a}\right)} *_t  {\rm vp}\frac{\exp(itx)}{\sinh\left(\frac{\pi(k-t)}{2a}\right)} \right](z).
\]
The linear change of the variable $k-t=u$ turns the above formula into
\be
\label{B3}
\left[{\rm vp}\frac{\exp(ixu)}{\sinh\left(\frac{\pi u }{2a}\right)} *_u  {\rm vp}\frac{\exp(-ixu)}{\sinh\left(\frac{\pi u}{2a}\right)} \right](2k-z).
\ee
Its imaginary part vanishes so finally the expression (\ref{B3}) consists of two components
\[
\left[{\rm vp}\frac{\cos(xu)}{\sinh\left(\frac{\pi u }{2a}\right)} *_u  {\rm vp}\frac{\cos(xu)}{\sinh\left(\frac{\pi u}{2a}\right)} \right](2k-z)+
\left[\frac{\sin(xu)}{\sinh\left(\frac{\pi u }{2a}\right)} *_u  \frac{\sin(xu)}{\sinh\left(\frac{\pi u}{2a}\right)} \right](2k-z).
\]
The second element is a convolution of functions and can be found by integration as
\be
\label{B3.1}
\int_{-\infty}^{+\infty}\frac{\sin(xu)}{\sinh\left(\frac{\pi u }{2a}\right)}  \frac{\sin(x[2k-z-u])}{\sinh\left(\frac{\pi [2k-z-u]}{2a}\right)} du.
\ee

On the other hand
\[
{\rm Re}\, \left\{{\rm vp}\frac{\exp(ixu)}{\sinh\left(\frac{\pi u }{2a}\right)} *_u  {\rm vp}\frac{\exp(ixu)}{\sinh\left(\frac{\pi u}{2a}\right)}\right\}=
{\rm vp}\frac{\cos(xu)}{\sinh\left(\frac{\pi u }{2a}\right)} *_u  {\rm vp}\frac{\cos(xu)}{\sinh\left(\frac{\pi u}{2a}\right)}-
\frac{\sin(xu)}{\sinh\left(\frac{\pi u }{2a}\right)} *_u  \frac{\sin(xu)}{\sinh\left(\frac{\pi u}{2a}\right)}.
\]
Therefore
\[
{\rm vp}\frac{\exp(ixu)}{\sinh\left(\frac{\pi u }{2a}\right)} *_u  {\rm vp}\frac{\exp(-ixu)}{\sinh\left(\frac{\pi u}{2a}\right)}=
{\rm Re}\, \left\{{\rm vp}\frac{\exp(ixu)}{\sinh\left(\frac{\pi u }{2a}\right)} *_u  {\rm vp}\frac{\exp(ixu)}{\sinh\left(\frac{\pi u}{2a}\right)}\right\}+2
\frac{\sin(xu)}{\sinh\left(\frac{\pi u }{2a}\right)} *_u  \frac{\sin(xu)}{\sinh\left(\frac{\pi u}{2a}\right)}.
\]
But
\be
\label{B31}
{\rm vp}\frac{\exp(ixu)}{\sinh\left(\frac{\pi u }{2a}\right)}= \tilde{\cal{F}}_t \left[ia \sqrt{\frac{2}{\pi}}\tanh\big(a(t+x)\big) \right](u).
\ee
Indeed,
\[
{\rm vp}\int_{-\infty}^{+\infty}\frac{\cos t u}{\sinh\left(\frac{\pi u }{2a}\right) }du =0\;\;\; {\rm and} \;\;\;
\int_{-\infty}^{+\infty}\frac{\sin tu}{\sinh\left(\frac{\pi u }{2a}\right) }du =a \tanh at
\]
(see \cite{gra}).
The observation (\ref{B31}) implies that
\[
{\rm vp}\frac{\exp(ixu)}{\sinh\left(\frac{\pi u }{2a}\right)}*_u {\rm vp}\frac{\exp(ixu)}{\sinh\left(\frac{\pi u }{2a}\right)}=
-2a^2 \sqrt{\frac{2}{\pi}} \tilde{\cal{F}}_t \big[\tanh^2 \big(a(t+x)\big)  \big](u).
\]
Moreover, the derivative
\be
\label{B4}
\frac{d \, \tanh \big(a(t+x)\big) }{dt}= a \left(1- \tanh^2 \big(a(t+x)\big)  \right).
\ee
Applying property (\ref{aa4}) to the expression (\ref{B4}) we obtain that
\[
\left({\rm vp}\frac{\exp(ixu)}{\sinh\left(\frac{\pi u }{2a}\right)}*_u {\rm vp}\frac{\exp(ixu)}{\sinh\left(\frac{\pi u }{2a}\right)} \right)(z)=-4a^2 \delta(z)+ 2z \cdot {\rm vp}\frac{\exp(ixz)}{\sinh\left(\frac{\pi z }{2a}\right)}.
\]
Thus
\[
\left({\rm vp}\frac{\exp(ixu)}{\sinh\left(\frac{\pi u }{2a}\right)} *_u  {\rm vp}\frac{\exp(-ixu)}{\sinh\left(\frac{\pi u}{2a}\right)}\right)(z)=
-4a^2 \delta(z)+ 2 \cos(xz) \cdot \frac{z}{\sinh\left(\frac{\pi z }{2a}\right)}+
\]
\[
  2 \left(\frac{\sin(xu)}{\sinh\left(\frac{\pi u }{2a}\right)} *_u  \frac{\sin(xu)}{\sinh\left(\frac{\pi u}{2a}\right)}\right)(z) .
\]
Finally
\[
W_{\frac{\hbar^2 k^2}{2m}}(x,p)=|A|^2\frac{k^2-a^2}{k^2+a^2}\delta(p+\hbar k )- \frac{2|A|^2 k }{\hbar (a^2+k^2)} \cos \left(2x \frac{k \hbar +p}{\hbar}\right) {\rm vp} \, \frac{1}{\sinh\left( \frac{ \pi(k \hbar +p)}{a \hbar}\right)}+
\]
\[
\frac{2|A|^2 }{\hbar^2 (a^2+k^2)}
  \cos \left(2 x\frac{k \hbar +p}{\hbar}\right)  \, \frac{k \hbar+p}{\sinh\left( \frac{ \pi(k \hbar +p)}{a \hbar}\right)}+
\]
\be
\label{B5}
\frac{|A|^2 }{\hbar (a^2+k^2)}
\int_{-\infty}^{+\infty}\frac{\sin(xu)}{\sinh\left(\frac{\pi u }{2a}\right)}  \frac{\sin\left(   \frac{x(2 k\hbar +2p-u)}{\hbar}   \right)}{\sinh\left( \frac{\pi(2 k\hbar +2p-u)}{2a \hbar} \right)} du.
\ee
The unusual relation that the momentum $p$ corresponds to the wave vector $-k$ results from the fact that our sign convention in the $*$ -- product is in agreement with Fedosov's papers \cite{6,7}.

Notice that for the state satisfying the condition $|k|=a$ the component containing the Dirac delta disappears.

\begin{figure}[H]\centering
\includegraphics[scale=1.0]{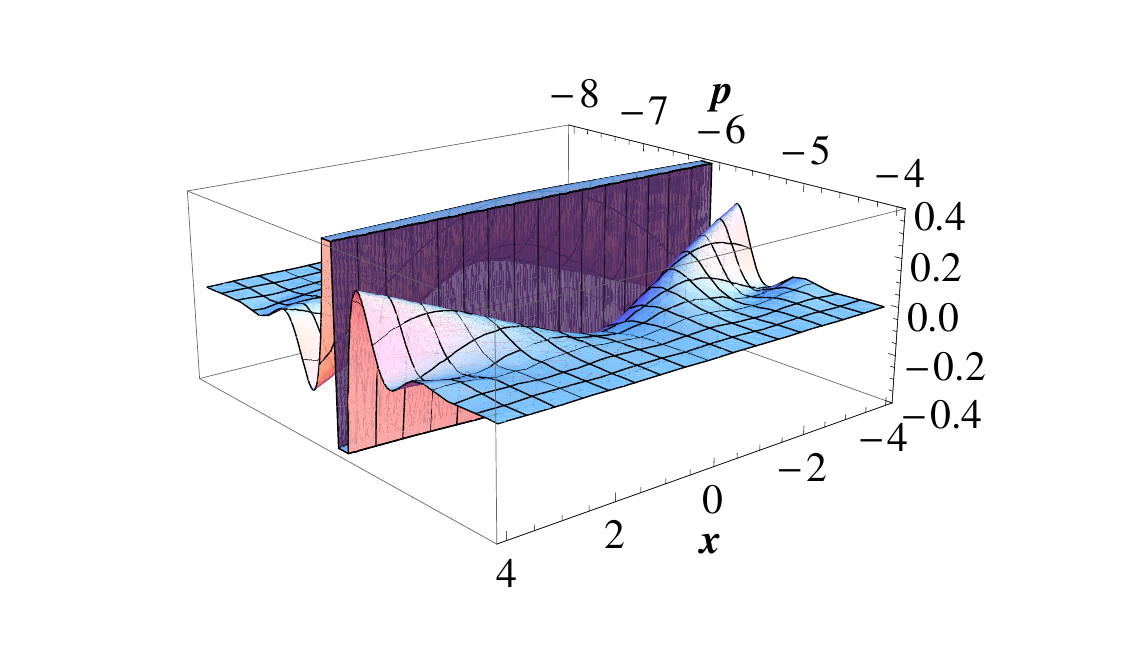}
\caption{The strict Wigner function  of the Poeschl -- Teller potential for $k=6,\, M=a=1$.}
\label{pic1.18}
\end{figure}


\end{document}